\documentclass[aps,preprint,groupedaddress]{revtex4}  

\usepackage{amssymb}
\usepackage{ulem}
\usepackage{amsmath}
\usepackage{amssymb}
\usepackage{graphicx}
\usepackage{subfigure}
\usepackage{color}
\usepackage{mathrsfs}
\usepackage[dvipsnames]{xcolor}
\definecolor{hg}{rgb}{0.0, 0.44, 0.0}

\usepackage[breaklinks=true,colorlinks=true]{hyperref}
\hypersetup{colorlinks=true,citecolor=blue,linkcolor=blue,urlcolor=blue}

\usepackage[utf8]{inputenc}

\begin{document}
\immediate\write16{<<WARNING: LINEDRAW macros work with emTeX-dvivers
                    and other drivers supporting emTeX \special's
                    (dviscr, dvihplj, dvidot, dvips, dviwin, etc.) >>}

\title{Vacuum state perturbed by an oscillon: composite configurations in $\phi^4$ model}

\author{Fabiano C. Simas$^{1,3}$ and Eduardo da Hora$^{2,3}$}

\email{fc.simas@ufma.br, carlos.hora@ufma.br }

\affiliation{$^{1}$ Departamento de Física, Universidade Federal do Maranhão (UFMA), Campus Universitário do Bacanga, 65085-580, São Luís, Maranhão, Brazil\\
$^{2}$ Coordenação do Curso de Bacharelado Interdisciplinar em Ciência e Tecnologia, Universidade Federal do Maranhão, 65080-805, São Luís, Maranhão, Brazil.\\
$^{3}$ Programa de Pós-Graduação em Física, Universidade Federal do Maranhão (UFMA), Campus Universitário do Bacanga, 65085-580, São Luís, Maranhão, Brazil
}

 
\begin{abstract}

We examine the evolution of a vacuum configuration when perturbed by an oscillon. We consider the $\phi^4$ scenario with a single scalar field only. For highly excited oscillons, we find that new composite solutions appear. They are formed by multiple antikink-kink pairs and a centered reminiscent oscillon. The overall process gives rise to a resonant structure whose pattern resembles that of a genuine kink-antikink scattering. We then approximate the initial oscillon by a lumplike profile constructed as a kink-antikink pair. We map the numerical results previously obtained. The developments proposed here may contribute to the understanding of the mechanism through which an oscillon decay reproduces the structure typically related to a kink-antikink collision. 

\end{abstract}


\maketitle

\section{Introduction}
\label{intro}


Topological structures are solutions to highly nonlinear models. Once nonlinearity is introduced through a symmetry-breaking potential, those structures are the result of a phase transition \cite{manton}. Under special circumstances, these solutions can also be obtained via a set of first-order differential equations. These equations appear from the minimization of the static total energy through the implementation of the Bogomol'nyi-Prasad-Sommerfield (BPS) technique \cite{prasom,bogo}.

In this context, localized structures have been studied in connection to a wide variety of physical systems, ranging from the subatomic to the cosmological scale \cite{dauxois,vacha}. Topological defects are particularly common in high-energy issues \cite{giblin}, optical communication \cite{mollgordon} and DNA \cite{dna}.

In a $(1+1)$-dimensional theory, the simplest topological solution is a kink. Its canonical version appears in a model with a single scalar field only. That solution exhibits a static profile with particle-like properties. Also, its spectrum against small excitations possesses both translational and vibrational modes. Energy exchange between these modes plays a fundamental role during kink-antikink scatterings. As a consequence, the resonant structure reveals a peculiar fractal pattern.

Traveling kinks have been widely investigated. In particular, the study of kink collisions revealed a surprisingly complex structure of bounces and annihilation. The outcome of the scattering process is also intimately influenced by the value of the initial velocity of the colliding structures. As a result, intricate resonance windows appear, see Refs. \cite{campbell,anninos,dorey}. Moreover, additional details on kink scattering in different scenarios, including the $\phi^6$, $\phi^8$ and deformed models, can be found in Refs.~\cite{takyi,limacarlos,henoli,weigel,gani1,dio1,simas1,dio2,dio3,diojoaza,alonso.2020,halava.2012,Lima.JHEP.2019,ivan,alonso.2021,adam.2019,adam.2021,adam.2019_1,joao.2023,joao.2021.2}.

At the same time, oscillons are spatially localized solutions with an oscillatory profile and a long lifetime. Their stability can not be assured due to the lack of a well-defined topological charge \cite{gleiser1,gleiser2}. These solutions naturally emerge from the collapse of bubbles \cite{cope}. They can also be formed in the context of a domain wall network \cite{hind}, and during the decay of a slightly perturbed spharelon \cite{manroman,manton2}. The relation between the dynamics of oscillons and sphalerons was also studied in Ref. \cite{olquromwere}, while a direct connection between sphalerons and the fractal pattern inherent to a kink-antikink collision was proposed in \cite{adamromawere}.

Research on oscillons in the $\phi^4$ model has become more active in recent years. Specifically, in Ref. \cite{alexeeva}, the authors have examined the existence and stability of the oscillon in the $\phi^4$ model using two different methods. In particular, they have verified an agreement between the numerical energy-frequency dependence and its variational counterpart. In the three-dimensional version of the $\phi^4$ model, a study was conducted in Ref. \cite{alexeeva2} to elucidate the resonant properties of oscillons through standing waves in a ball. Another work that has adopted a variational formulation is presented in Ref. \cite{alexeeva3}. In this case, the authors have identified the onset of oscillon instability. Additional studies on oscillons are available in Refs. \cite{dorey2,baraprl,nagy,hirama,hong}. Also, it is now recognized that oscillons play an important role in gravitational effects, black holes, inflationary reheating and dark matter \cite{zhang,nazari,amin,olle}.

In addition, Ref. \cite{roman_osc} has revealed the creation of kink-antikink pairs via the collision between two identical wave trains in the $\phi^4$ model. They have identified that the pair generation consists of three distinct phases. Also, they have observed that the resulting fractal structure is determined by the amplitude and number of perturbations.

Most recently, Ref. \cite{osc_weres} has investigated the fractal pattern that appears from the decay of an excited oscillon in a model with a symmetric vacuum only. That effective model does not support static solutions with particle-like properties. However, the resonant structure, which is based on the energy exchange between the internal modes, is realized. Additional investigations on the pair production can also be found in Refs. \cite{vacha1,dele,dele1,carpe}.

The aim of the present manuscript is to contribute with such a research. Here, in the context of the $\phi^4$ model, we consider the evolution of a vacuum state when perturbed by an oscillon. The initial oscillon is controlled by two parameters, $A_0$ and $\sigma$. They represent the amplitude and width of that structure, respectively. As we demonstrate below, the value of $A_0$ determines the final result of field decay. In particular, as $A_0$ approaches 1 (highly excited oscillon), composite configurations emerge. These profiles are formed by multiple kink-antikink pairs and centered reminiscent fluctuations. We also show that the overall process gives rise to a resonant structure whose pattern mimics that of a genuine kink-antikink scattering.

Our manuscript is organized as follows: in Sec. \ref{secII}, we introduce the model and define the initial condition. We then study the field evolution for different values of the amplitude, with width fixed. We depict the numerical results and comment on their main aspects. We also plot the corresponding resonant structures. In Section \ref{secIII}, we approximate the initial oscillon by a lumplike profile constructed as a kink-antikink pair. We then map some of the results introduced in Sec. \ref{secII}. Finally, Section \ref{secIV} brings a brief summary and our perspectives regarding future research.

\section{The model and its results} \label{secII}

We consider a scenario with one real scalar field that self-interacts via the potential $V(\phi)$. The Lagrangian density is
\begin{eqnarray}
    {\mathcal{L}} = \frac12 \partial_{\mu} \phi \partial^{\mu} \phi - V(\phi).\label{model}
\end{eqnarray}
Here, field, coordinates and coupling constants are assumed to be dimensionless (can be achieved via an appropriate mass rescaling). The Greek index $\mu$ runs from 0 to 1. In addition, the metric of the Minkowski spacetime is assumed to be $(+-)$.

The potential $V(\phi)$ determines the vacuum structure of the corresponding theory. In this manuscript, we adopt the usual $\phi^4$ one, i.e.
\begin{eqnarray}
    V(\phi) = \frac12(1-\phi^2)^2.\label{potential}
\end{eqnarray}
This potential has many applications in different scenarios. In condensed matter physics, for example, a one-dimensional Ginzburg-Landau theory has been proposed to describe the phase transition in shape memory alloys \cite{falk}. Also, in Ref. \cite{grirub}, the authors studied how the temperature influences the formation of kink-antikink pairs.

The $\phi^4$ model is known to possess analytical time-independent solutions with nontrivial topology. These kink profiles connect the fundamental states predicted by the potential (\ref{potential}). In this sense, kinks and antikinks are related via the $Z_2$ symmetry inherent to the model (\ref{model}). These structures have the same energy, but topological charges with opposite signs. In addition, their spectrum against small perturbations possesses both translational and vibrational modes. As a consequence, the evolution of a kink-antikink collision is affected by these modes via the {\it{energy transfer mechanism}}. Two-bounce windows then appear and a typical fractal structure emerges.

Instead of colliding kinks, we now investigate the evolution of a vacuum state when perturbed by an oscillon. With such a purpose in mind, we consider the initial condition
\begin{eqnarray}
    \phi(x,0) = 1 + A_0 e^{-x^2/\sigma},\label{ic}
\end{eqnarray}
where $A_0$ and $\sigma$ represent the amplitude and width of the oscillon, respectively. Here, the Gaussian perturbation was applied to the vacuum $\phi_0 = +1$. Naturally, depending on the values of $A_0$ and $\sigma$, the resulting oscillons may produce different final states.

The initial configuration (\ref{ic}) is comparable to the one used in Ref. \cite{osc_weres}. In that work, the authors have studied the behavior of a symmetric vacuum when submitted to an oscillon. However, as their model does not support static particle-like configurations, the authors have approximated the potential by the sine-Gordon one. As a consequence, they have successfully mapped the evolution of $\phi$ in terms of the analytical breather solution.

The present manuscript offers an alternative example. The point is that the potential (\ref{potential}) does support static kinks with particle-like properties. These solutions can be used to mimic the initial configuration (\ref{ic}). The resonant structure related to the oscillon decay can then be compared with that of a genuine kink-antikink interaction. The conclusion is that a kink-antikink pair can be used to preview qualitatively the evolution of the fundamental state when perturbed by the oscillon.

In what follows, we investigate the evolution of $\phi(x,t)$ in the full space for different values of $A_0$ and fixed $\sigma$. We have solved the equation of motion in a box $-z_{max}<x<z_{max}$, with $z_{max}=200$ and a space step $\delta x=0.05$. The partial derivatives in $x$ were approximated using the five-point stencil. The resulting set of equations was integrated using fifth-order Runge-Kutta algorithm with adaptive step size and periodic boundary conditions. At this first moment, we are interested in small perturbations around the vacuum state. We therefore consider very small values of the amplitude $A_0$. 

\begin{figure*}[!ht]
\begin{center}
  \centering
    \subfigure[]{\includegraphics[width=0.47
        \textwidth]{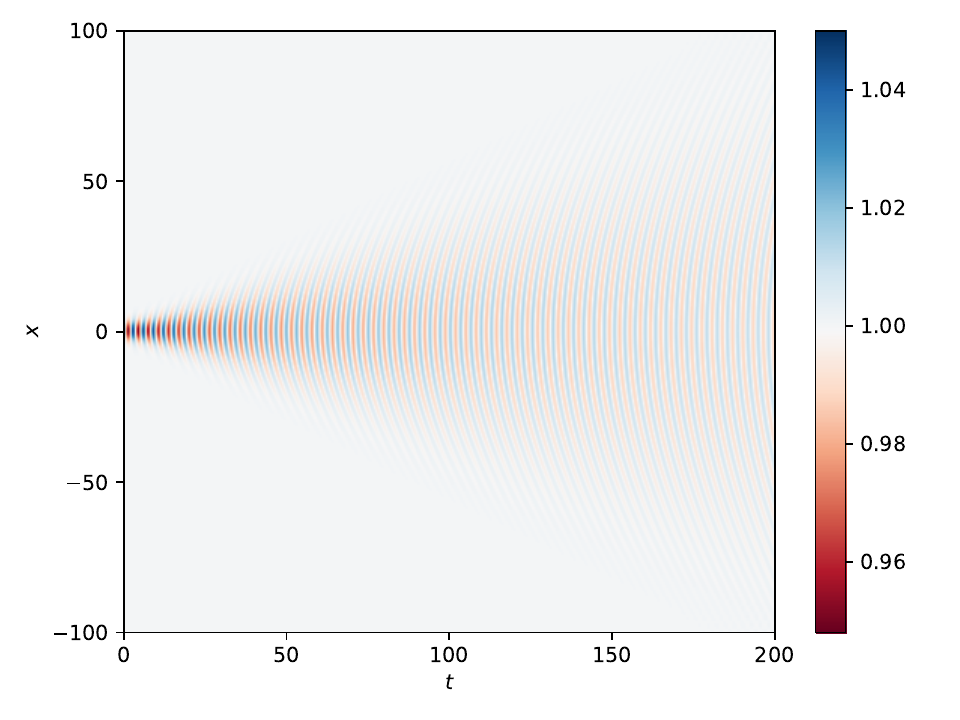}\label{s10A005}}
    \subfigure[]{\includegraphics[width=0.47
        \textwidth]{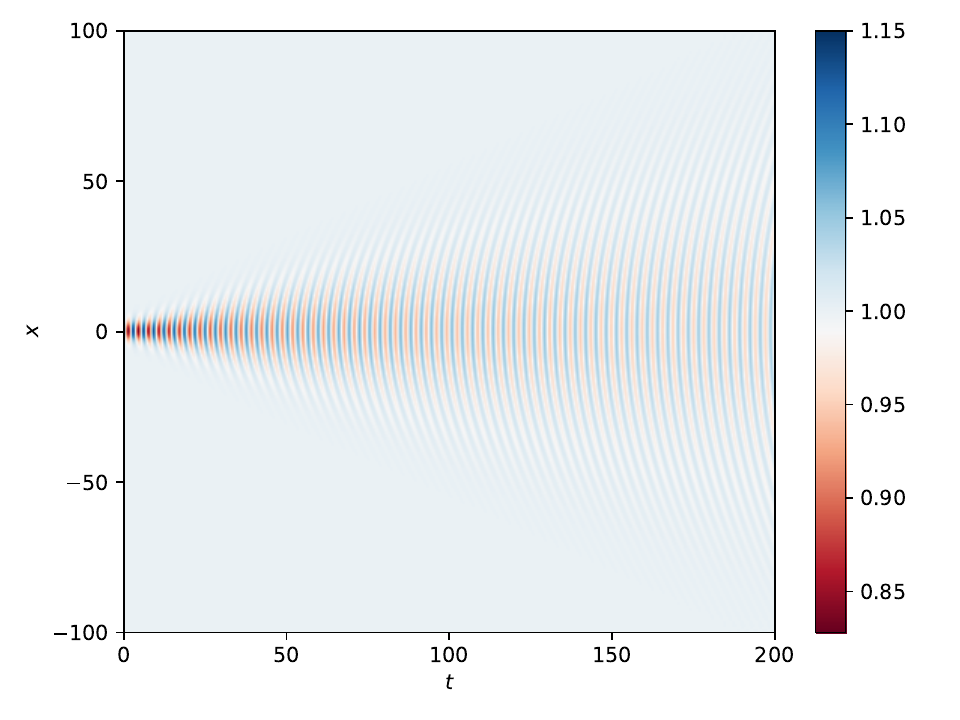}\label{s10A015}}
    \subfigure[]{\includegraphics[width=0.47
        \textwidth]{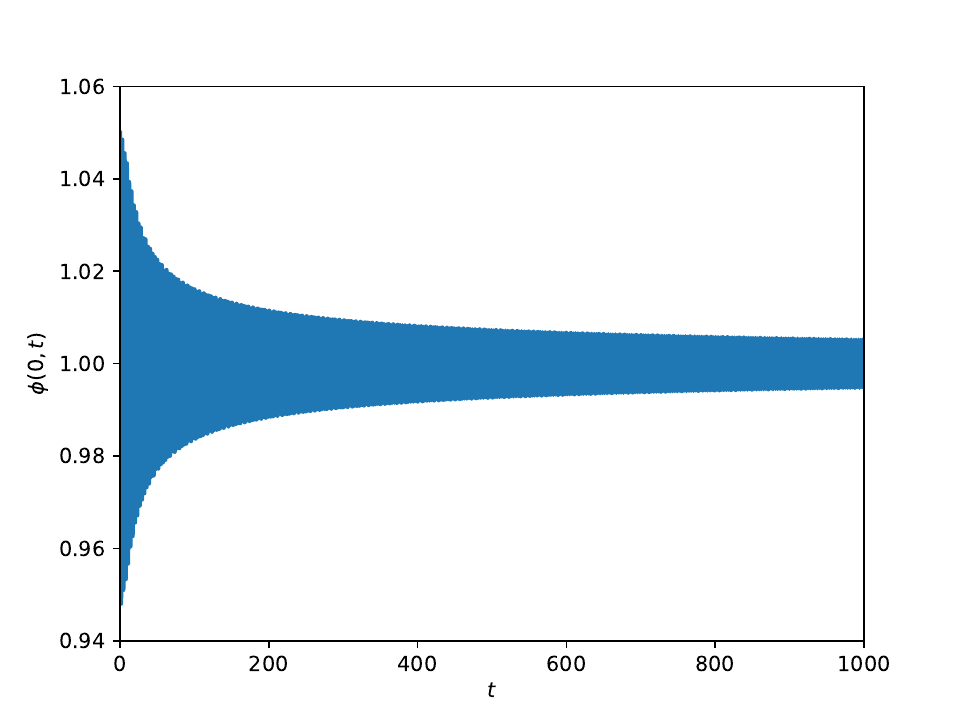}\label{s10A005_1}}
    \subfigure[]{\includegraphics[width=0.47
        \textwidth]{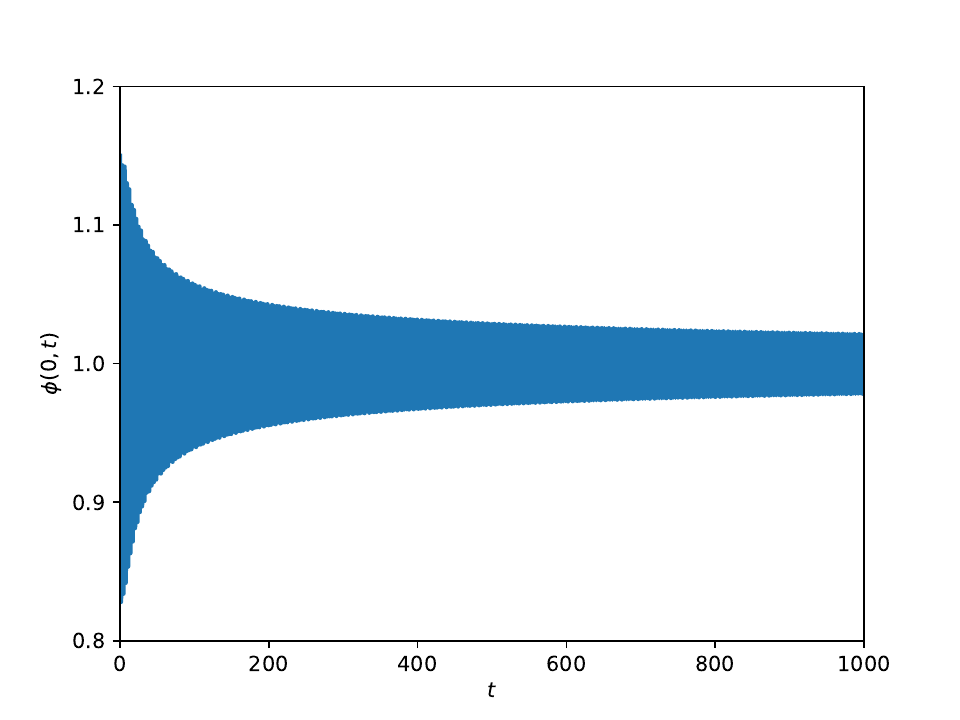}\label{s10A015}}
  \caption{Evolution of $\phi(x,t)$ in the full space (top) and the field profile at the center of mass (bottom) for $A_0=0.05$ (left) and $0.15$ (right). Here, $\sigma=10$. The reminiscent oscillon annihilates and the field returns to its fundamental state.}
  \label{fig_s10_A01}
\end{center}
\end{figure*}

Figure \ref{fig_s10_A01} (top) shows the evolution of $\phi(x,t)$ for $\sigma=10$, with $A_0=0.05$ (left) and $0.15$ (right). The original oscillon produces a small amplitude reminiscent one that fluctuates around the vacuum state. In the sequence, the reminiscent oscillon decays and annihilates. This annihilation is due to the emission of radiation, an effect related to the nonintegrable nature of the $\phi^4$ model. The field profile at the center of mass as a function of $t$ is also displayed in Fig. \ref{fig_s10_A01} (bottom). It reveals how the amplitude of the reminiscent fluctuations decreases.

The conclusion is that the scalar field tends to return to its fundamental state when perturbed by an oscillon with small amplitude. In other words, the perturbation are not intense enough to remove the field from its vacuum value and then generate a nontrivial configuration. Remarkably, unlike what is seen in Ref. \cite{osc_weres}, the perturbation here does not decompose into {\textit{constituent}} structures.

We also consider the case of not so small amplitudes. Such a range was not considered in Ref. \cite{osc_weres}. Here, as the amplitude $A_0$ increases, the initial oscillons become more excited. As a consequence, the field decays into new interesting configurations. In particular, when $A_0$ approaches 1, composite solutions appear. These configurations are formed by multiple structures, including kink-antikink pairs and reminiscent oscillons. As we have pointed out, these composite configurations find no direct correspondence with the previous results in \cite{osc_weres}. In this sense, the evolution of a vacuum state when submitted to such an initial oscillon represents an additional novelty introduced by the present investigation.

\begin{figure*}[!ht]
\begin{center}
  \centering
  \subfigure[]{\includegraphics[width=0.47
    \textwidth]{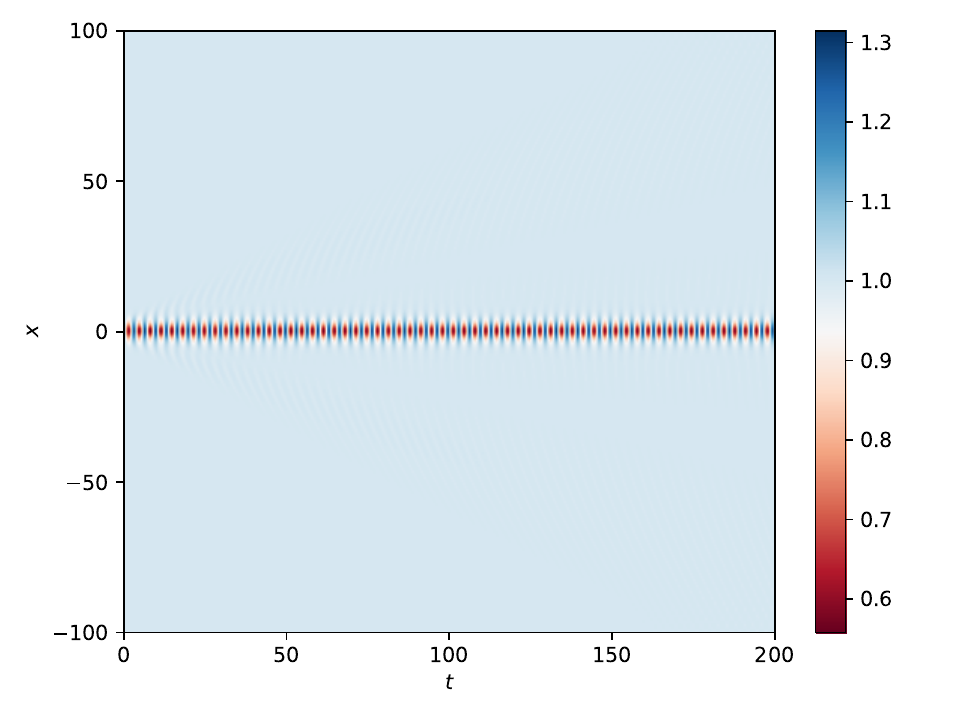}\label{s10A03}}
  \subfigure[]{\includegraphics[width=0.47
    \textwidth]{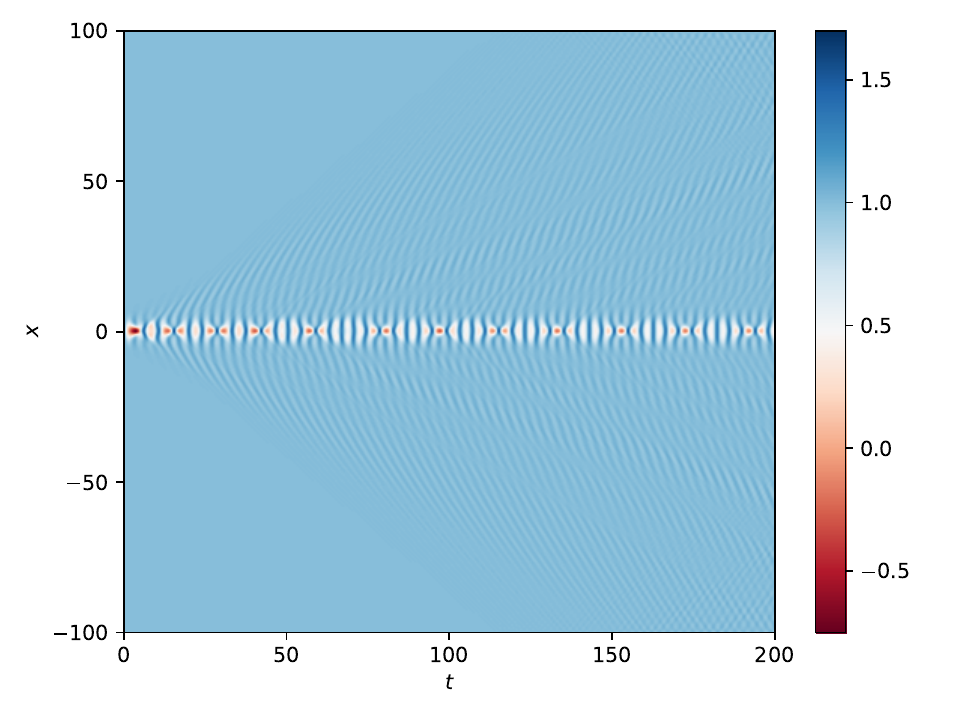}\label{s10A05}}
  \subfigure[]{\includegraphics[width=0.47
    \textwidth]{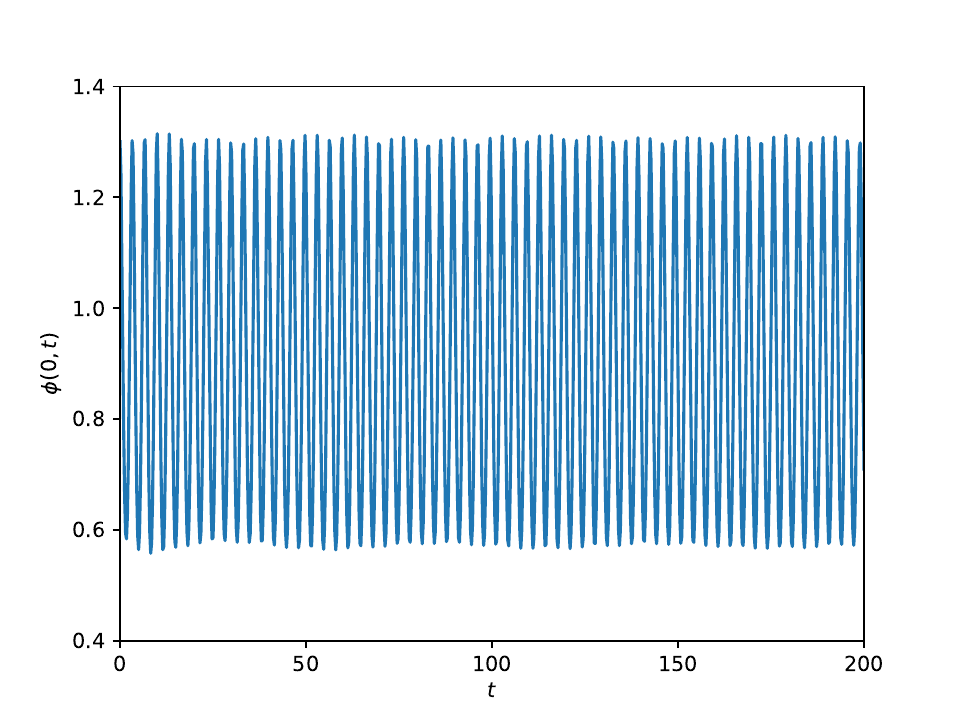}\label{s10A003_1}}
  \subfigure[]{\includegraphics[width=0.47
    \textwidth]{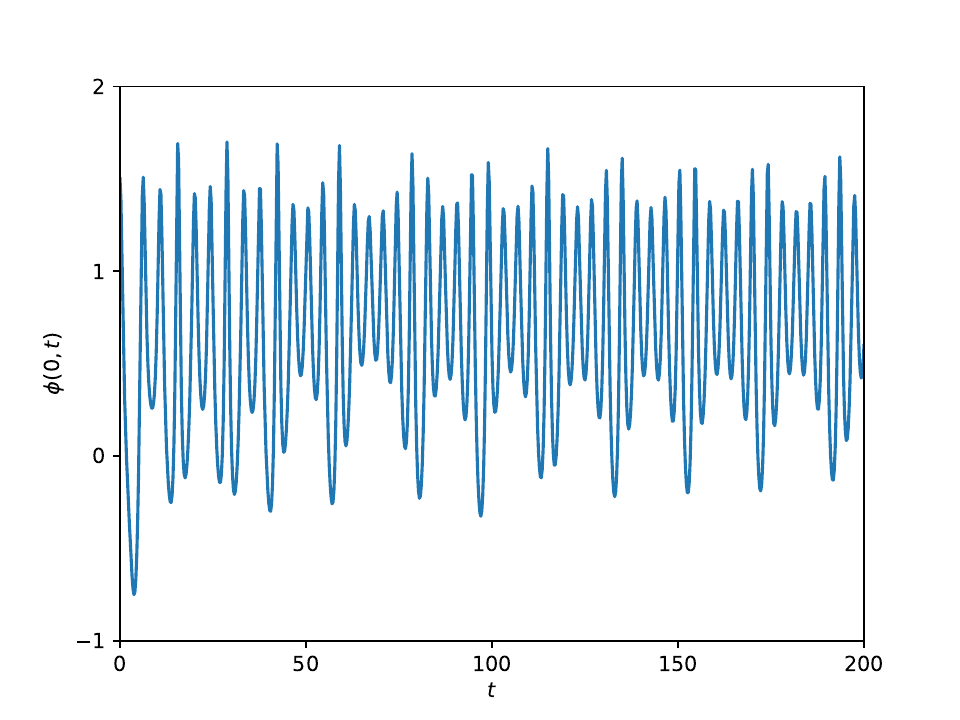}\label{s10A005_1}}
  \caption{Evolution of $\phi(x,t)$ in the full space (top) and the field profile at the center of mass (bottom) for $A_0=0.30$ (left) and $0.50$ (right). Here, $\sigma=10$.}
  \label{fig_s10_A_1}
\end{center}
\end{figure*}

We have solved the numerical problem for additional values of $A_0$, again with fixed $\sigma$. The corresponding solutions appear in the Fig. \ref{fig_s10_A_1} for intermediary $A_0$. The resulting solutions present a transitory behavior. In particular, for $A_0=0.30$, the perturbation produces a reminiscent oscillon that annihilates after an extremely long time. Its oscillations possess an {\textit{almost harmonic}} profile, and its annihilation induces the field to finally return to its vacuum configuration. Furthermore, the solution for $A_0=0.50$ reveals a reminiscent {\it{anharmonic}} oscillon, whose formation might be influenced by a more intricate dynamical process. At this amplitude, the oscillon solution begins to be perturbed. The perturbation increases for larger values of $A_0$. As a result, it leads to the formation of an antikink-kink pair.

\begin{figure*}[!ht]
\begin{center}
  \centering
  \subfigure[]{\includegraphics[width=0.3
    \textwidth]{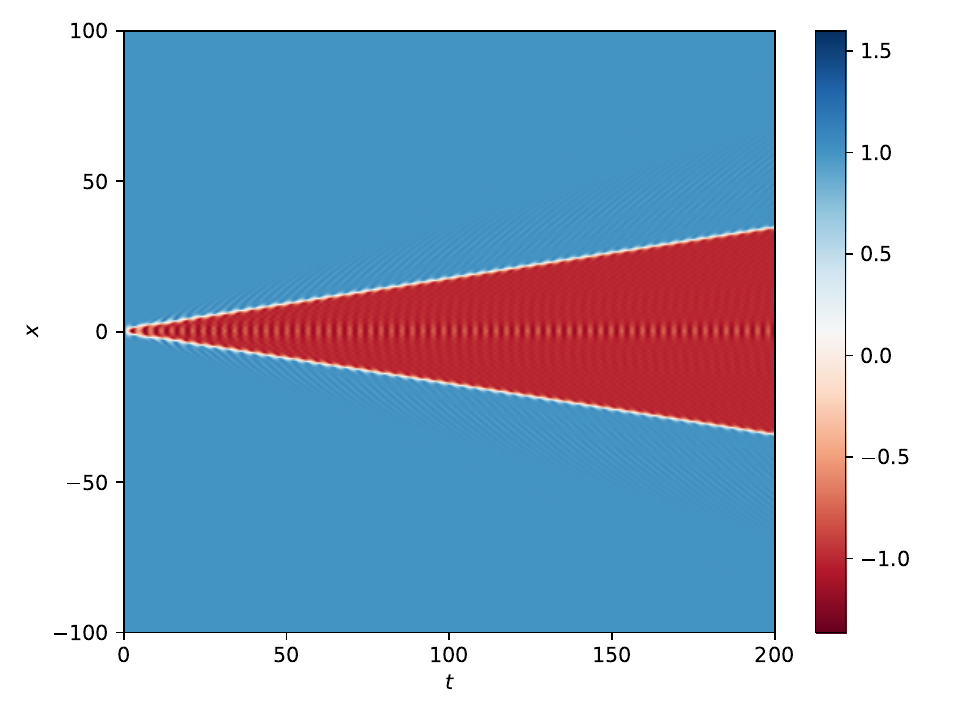}\label{s10A06}}
  \subfigure[]{\includegraphics[width=0.3
    \textwidth]{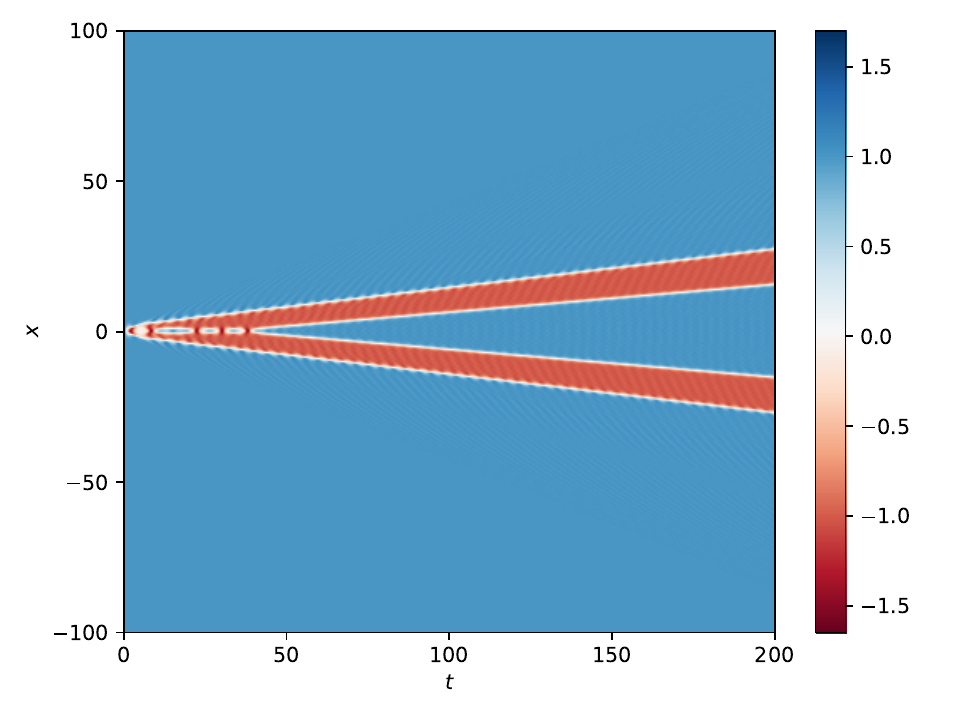}\label{s10A07}}
  \subfigure[]{\includegraphics[width=0.3
    \textwidth]{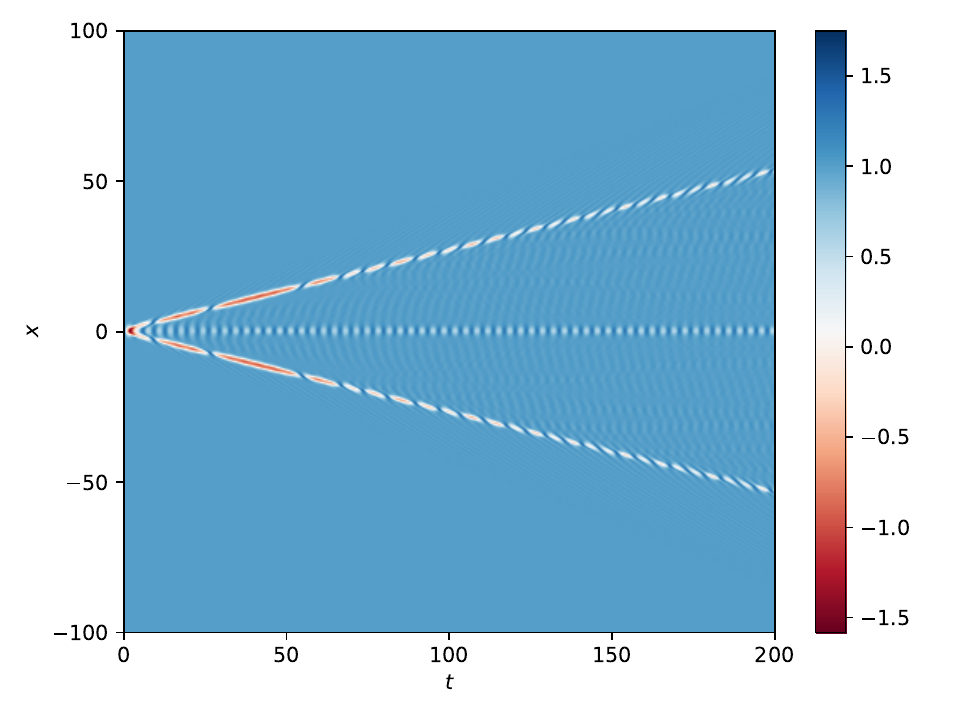}\label{s10A075}}
  \subfigure[]{\includegraphics[width=0.3
    \textwidth]{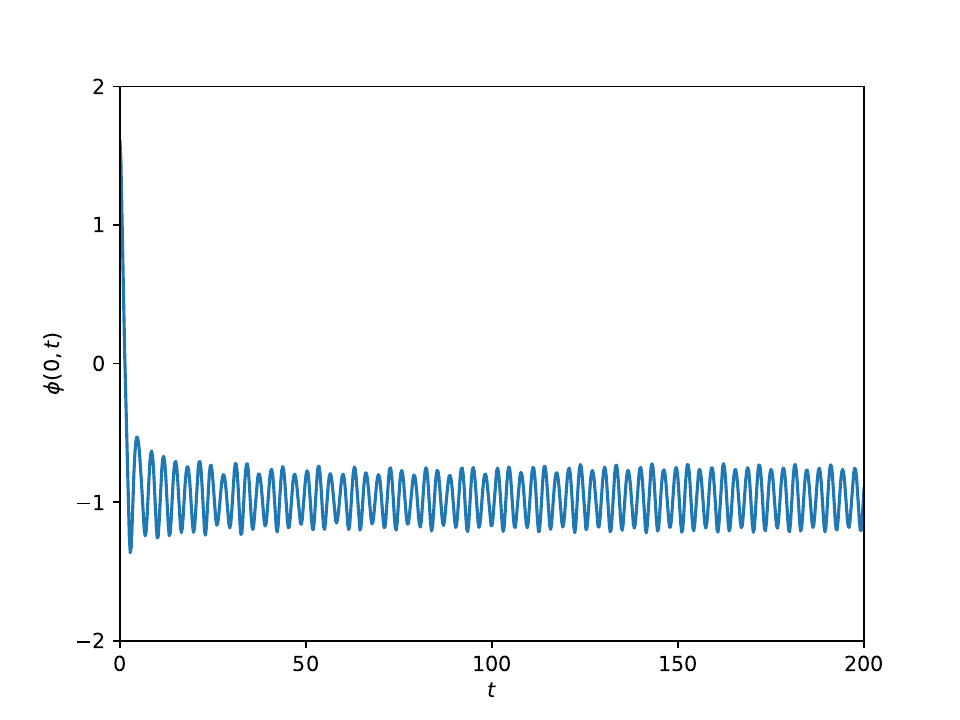}\label{s10A006_1}}
  \subfigure[]{\includegraphics[width=0.3
    \textwidth]{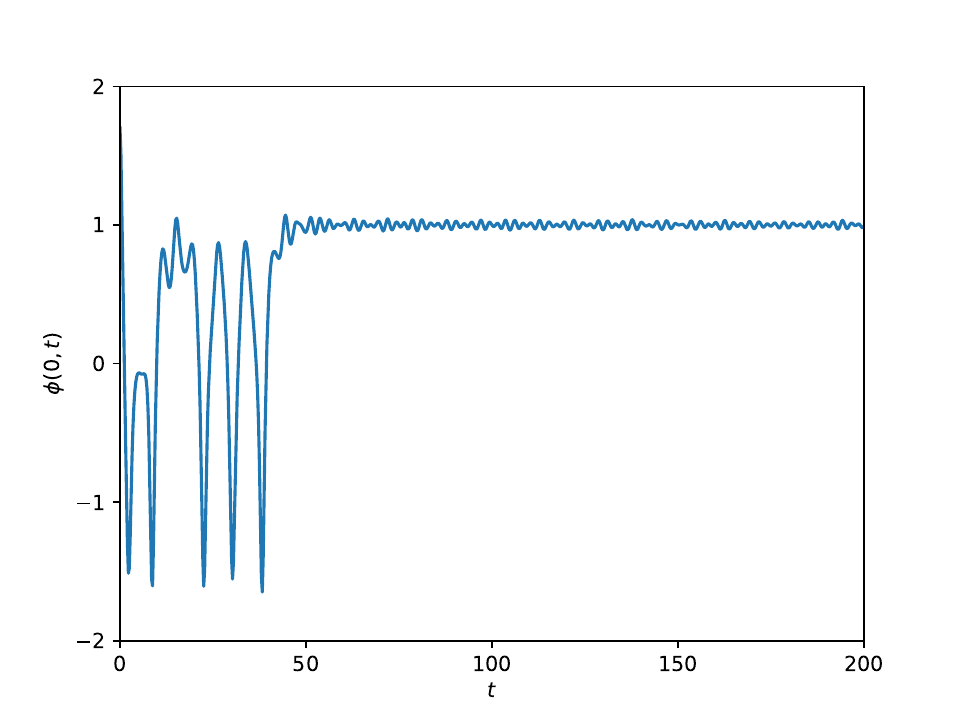}\label{s10A007_1}}
  \subfigure[]{\includegraphics[width=0.3
    \textwidth]{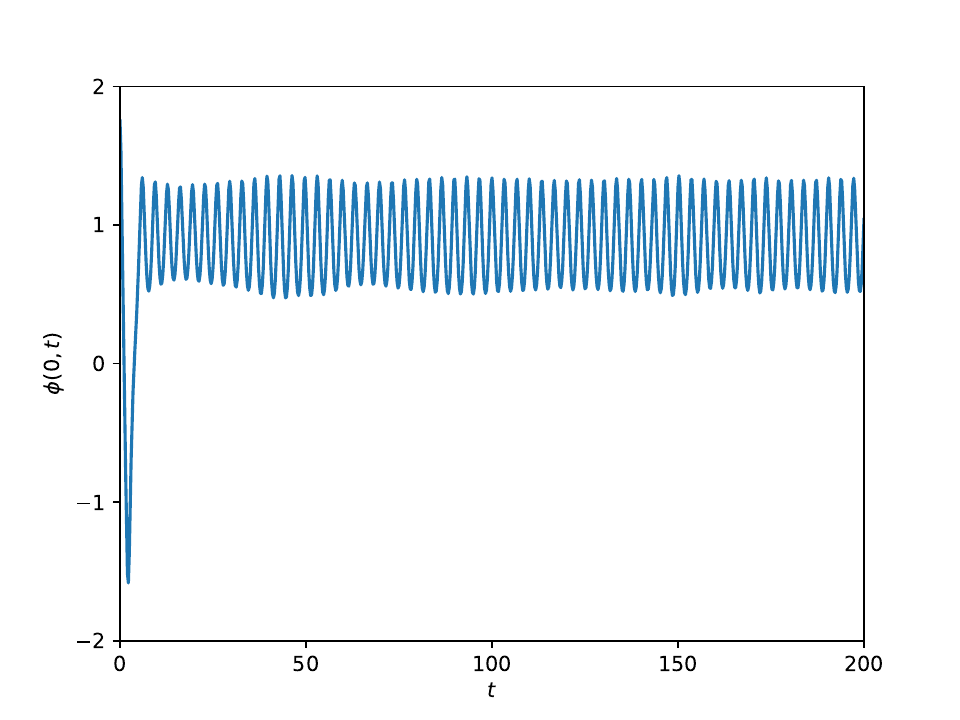}\label{s10A0075_1}}
  \caption{Evolution of $\phi(x,t)$ in the full space (top) and the field profile at the center of mass (bottom) for $A_0=0.60$ (left), $0.70$ (center) and $0.75$ (right). Here, $\sigma=10$. The initial oscillon rapidly decays into kink-antikink pairs.}
  \label{fig_s10_A_2}
\end{center}
\end{figure*}

The observation above is corroborated by the solutions that appear in Figure \ref{fig_s10_A_2}. Here, for $A_0=0.60$, the original perturbation rapidly decays into an antikink-kink pair with a centered reminiscent oscillon around the vacuum state $\phi_0=-1$. In addition, when $A_0=0.70$, two antikink-kink pairs emerge. Additional numerical simulations indicate that, as $A_0$ increases, the original oscillon decays and then creates antikink-kink pairs followed by the escape of the radiation.

Similar phenomenon was studied in Ref. \cite{fab_lump}, where the scattering of two lumps was verified to produce kink-antikink pairs. Multiple antikink-kink pairs were also obtained in a double sine-Gordon model \cite{fab_DsG}. Kink-antikink configurations also emerge from the collision between two identical wave trains, see Ref. \cite{roman_osc}.

\begin{figure*}[!ht]
\begin{center}
  \centering
    \subfigure[]{\includegraphics[width=0.47
    \textwidth]{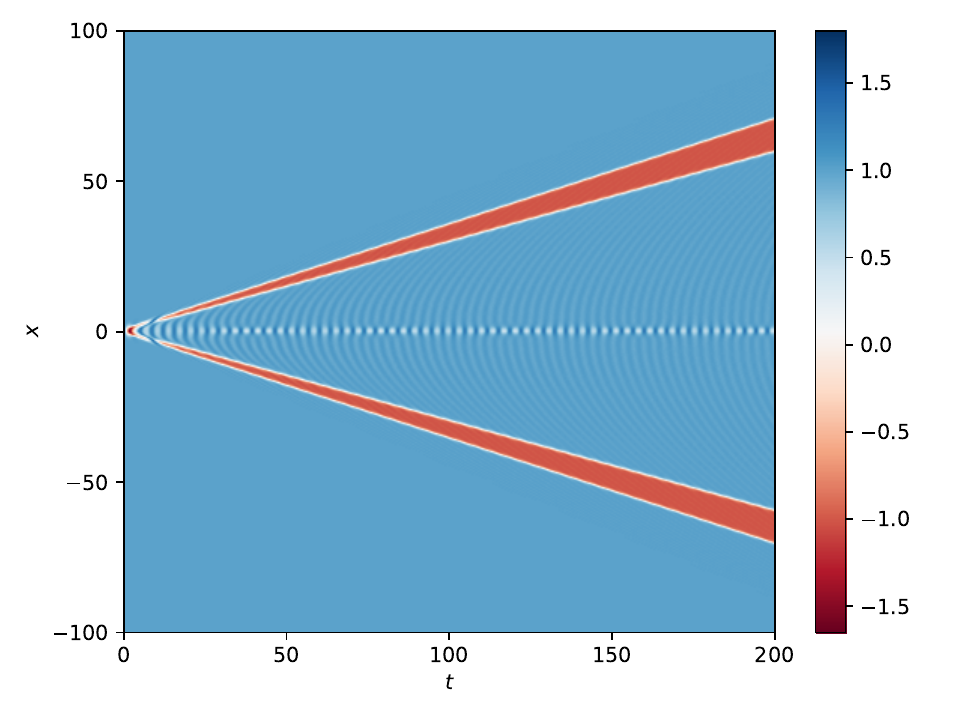}\label{s10A08}}
    \subfigure[]{\includegraphics[width=0.47
    \textwidth]{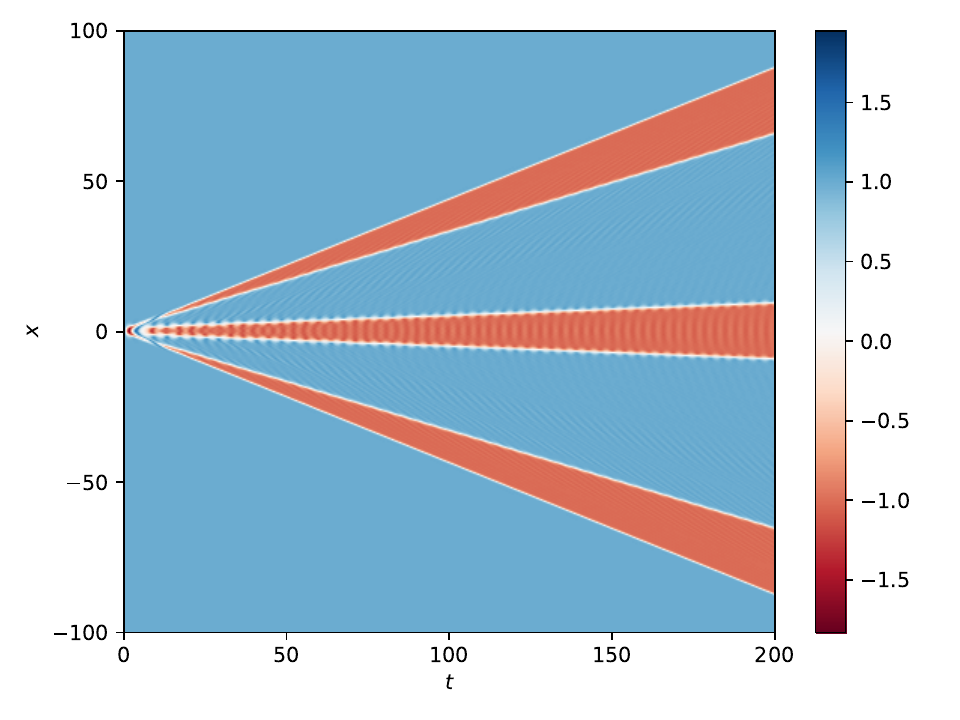}\label{s10A095}}
    \subfigure[]{\includegraphics[width=0.47
    \textwidth]{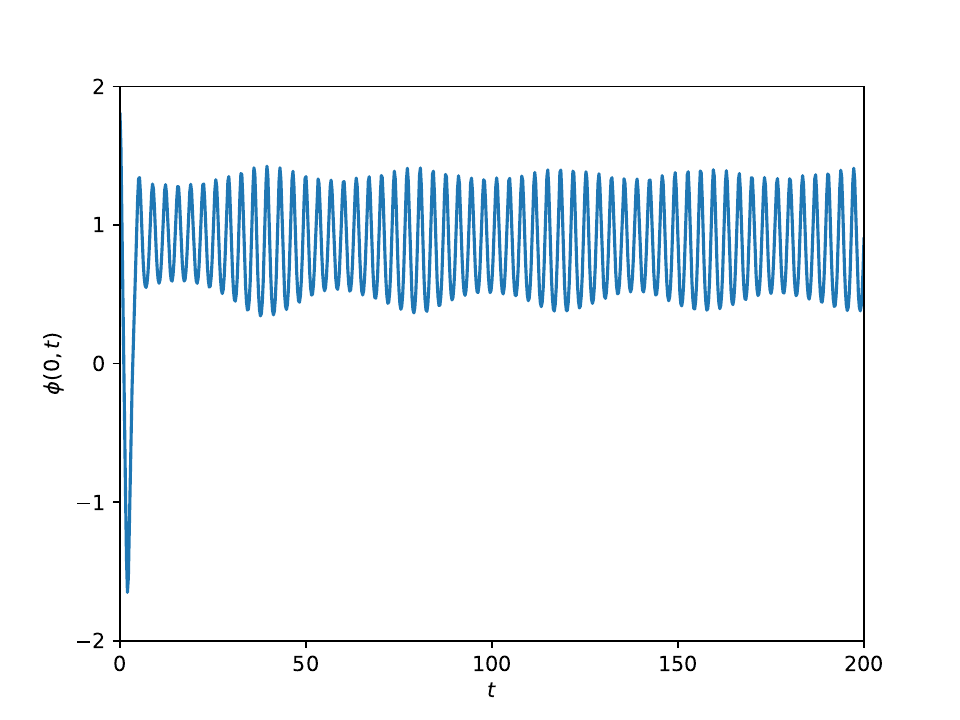}\label{s10A008_1}}
    \subfigure[]{\includegraphics[width=0.47
    \textwidth]{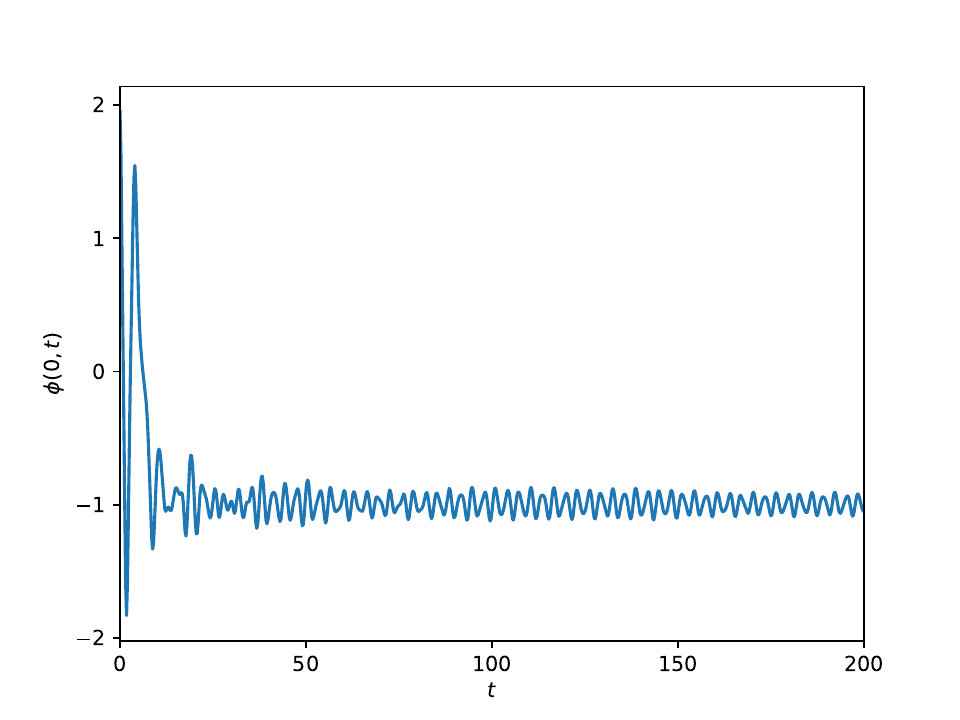}\label{s10A0095_1}}
  \caption{Evolution of $\phi(x,t)$ in the full space (top) and the field profile at the center of mass (bottom) for $A_0=0.80$ (left) and $0.95$ (right). Here, $\sigma=10$. The initial oscillon generates antikink-kink pairs. In particular, three non-bouncing pairs emerge for $A_0=0.95$.}
  \label{fig_s10_A_3}
\end{center}
\end{figure*}

We now return to our solution for $A_0=0.70$. In this case, we highlight that, once the evolution begins, centered fluctuations intermediate a ``collision" between the inner structures. The corresponding pattern mimics the one typically related to a four-bounce window, see the profile for $\phi(0,t)$. In particular, between the first two bounces, the scalar field performs two complete oscillations. This can be understood as a manifestation of a mechanism similar to the energy exchange one. Moreover, when $A_0=0.75$, the perturbation produces a pair of bions and a centered oscillon around $\phi_0=+1$.

\begin{figure*}[!ht]
\begin{center}
  \centering
    \subfigure[]{\includegraphics[width=0.495
    \textwidth]{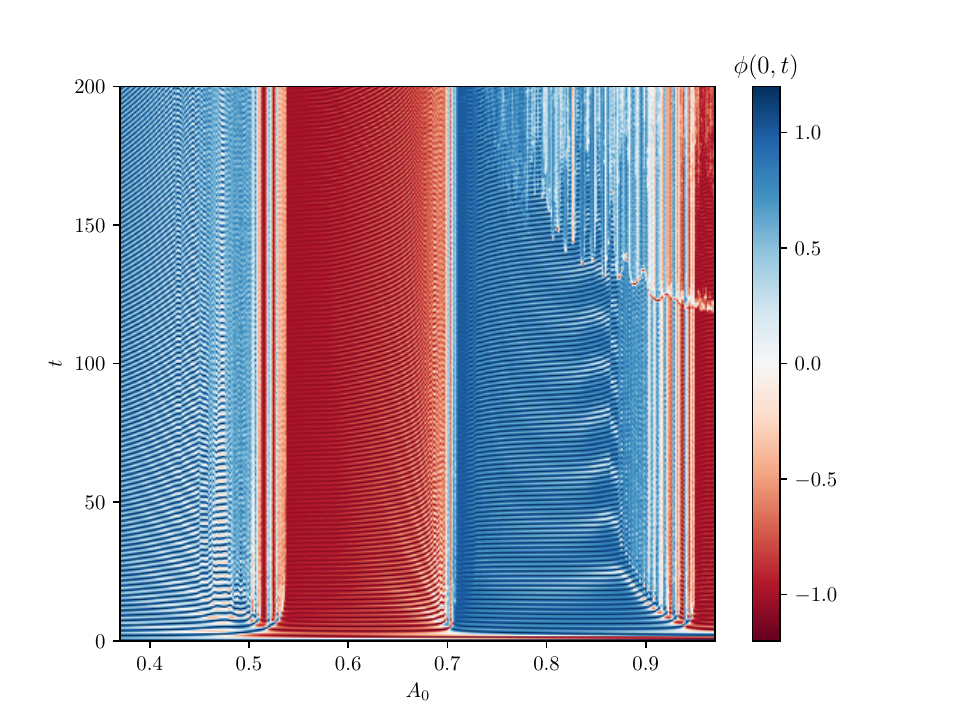}\label{tAs10_1}}
    \subfigure[]{\includegraphics[width=0.495
    \textwidth]{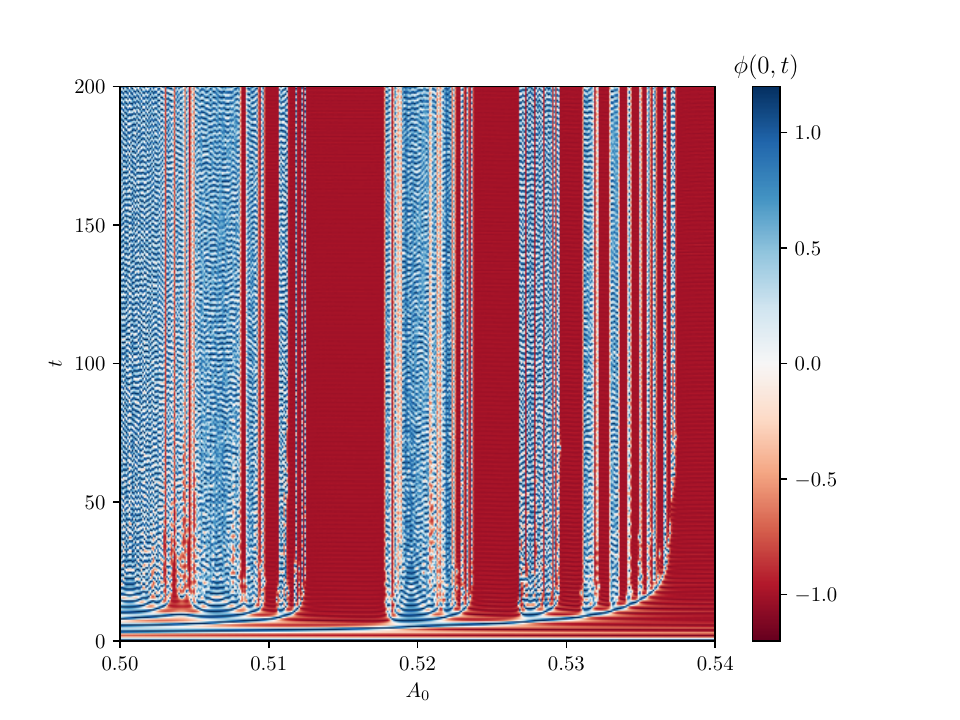}\label{tAs10_3}}
  \caption{The resonant structure related to the evolution of the vacuum state $\phi_0=+1$ when perturbed by an oscillon. Blue regions represent field oscillations around that state. Red portions mean the emergence of a centered antikink-kink pair. The structure resembles that of a kink-antikink collision. Here, we have fixed $\sigma=10$.}
  \label{times_A_sigma_10}
\end{center}
\end{figure*}

The final configurations obtained for high amplitudes are shown in Figure \ref{fig_s10_A_3}. In general, the initial condition oscillates and then generates two antikink-kink pairs. These pairs propagate without considerable radiation escape. At the same time, reminiscent oscillons now travel around while emitting a certain quantity of radiation.

Interestingly, there are ranges for which the initial perturbation also produces two oscillons that escape to infinity. In particular, two antikink-kink pairs appear for $A_0=0.80$. However, there are now two notable differences in comparison to the previous case: long-lived fluctuations around $\phi_0=+1$ take place at the center of mass, and the arising antikink-kink pairs do not bounce. Note that three non-bouncing pairs emerge for $A_0=0.95$.

We also consider the resonant structure related to the evolution of $\phi(x,t)$. Figure \ref{times_A_sigma_10} summarizes the main behaviors involving the field decay, as well as the appearance of antikink-kink pairs. These plots show the temporal evolution of the field value at the center of mass for fixed $\sigma=10$ and different $A_0$. It is possible to identify well-defined segments that alternate between blue and red colors. The blue regions represent fluctuations around $\phi_0=+1$, while the red portions mean the emergence of a centered antikink-kink pair. For instance, small values of $A_0$ lead to either an annihilation of the original perturbation or the generation of a long-lived reminiscent oscillon. On the other hand, when $0.5<A_0<0.7$, there are stripes in red that represent a single antikink-kink pair.

\begin{figure*}[!ht]
\begin{center}
  \centering
    \subfigure[]{\includegraphics[width=0.32
    \textwidth]{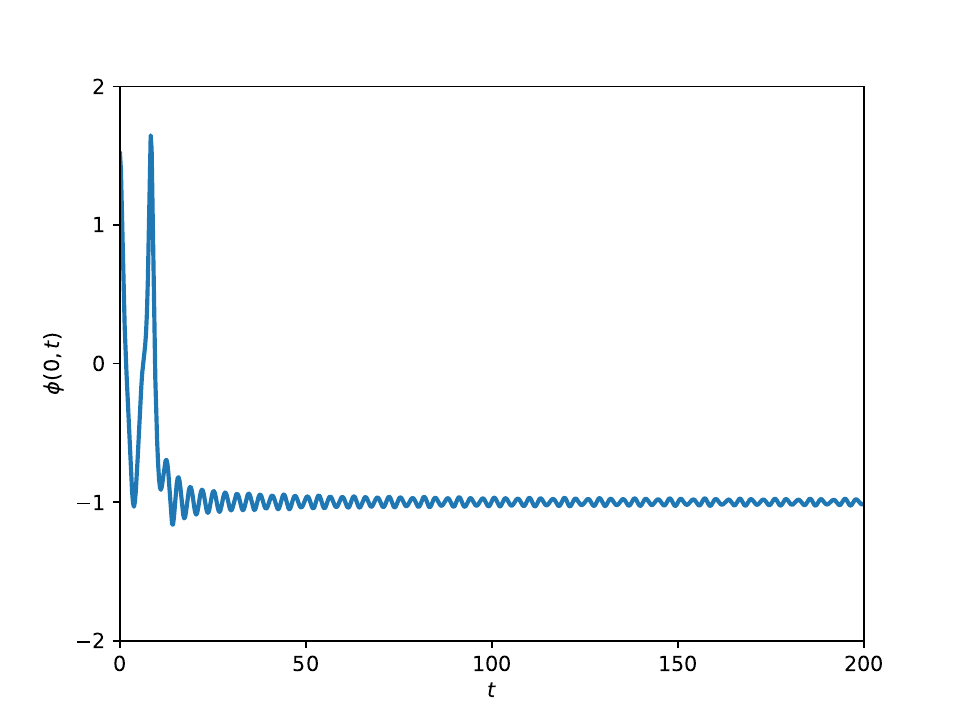}\label{s10A0515}}
    \subfigure[]{\includegraphics[width=0.32
    \textwidth]{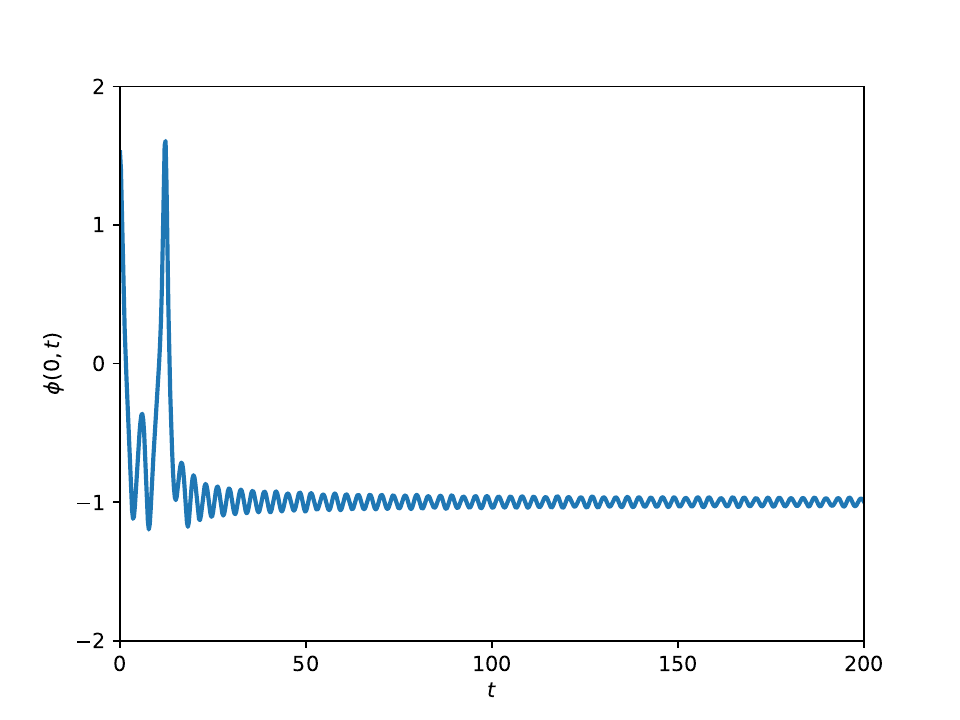}\label{s10A0525}}
    \subfigure[]{\includegraphics[width=0.32
    \textwidth]{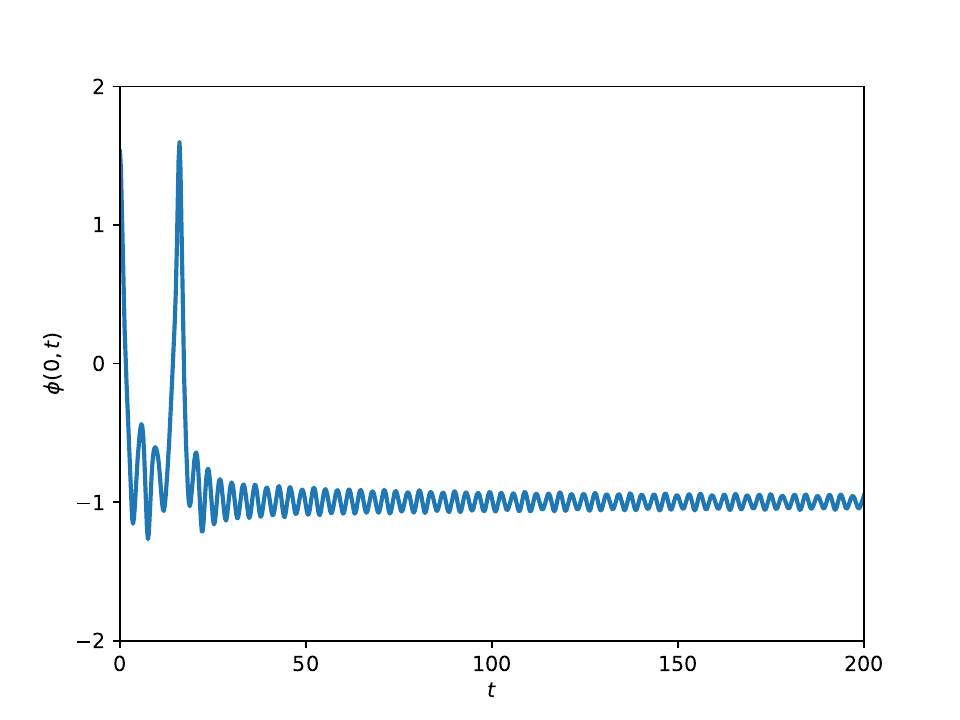}\label{s10A053}}
  \caption{The field profile at the center of mass for $A_0=0.515$ (left), $A_0=0.525$ (center) and $A_0=0.53$ (right). Here, we have fixed $\sigma=10$.}
  \label{fig_s10_A_0515}
\end{center}
\end{figure*}

Surprisingly, within this range of $A_0$, we discover a structure that resembles the outcome of a kink-antikink collision \cite{campbell,dorey}. To highlight its details, we have included a zoomed-in view, see Fig.~\ref{tAs10_3}. In addition, a similar pattern may also be found in Ref. \cite{osc_weres}, which addressed the decay of the oscillon in a model with only one vacuum. It is important to make clear that, concerning kink-antikink scattering, the appearance of a resonant structure is typically associated with the number of bounces performed by the pair. In our case, the plot illustrates an alternation between the formation of long-lived oscillons and the emergence of an antikink-kink pair.

The scattering of kinks in the $\phi^4$ model shows a two-bounce window structure. The order of the corresponding window is then determined by the number of oscillations $m$ between the bounces. Therefore, the first resonance window corresponds to $m=1$, see Ref. \cite{anninos}. These resonant peaks accumulate and decrease in thickness as the initial velocity increases. In our study, a similar process is used to determine the order of each resonant window. The behavior of the field at the center of mass for specific values of $A_0$ is depicted in Fig. \ref{fig_s10_A_0515}. The results reveal a bounce near the origin, followed by another peak after a certain time interval, with or without the presence of oscillations between these bounces. For instance, a zero-order window ($m=0$) is observed for $\sigma=10$ and $A_0=0.515$. Furthermore, for $A_0=0.525$ and $A_0=0.53$, the first ($m=1$) and second ($m=2$) resonant windows, respectively, are represented by one and two oscillations between the bounces.

\begin{figure*}[!ht]
\begin{center}
  \centering
    \subfigure[]{\includegraphics[{angle=0,width=8cm,height=6cm}]{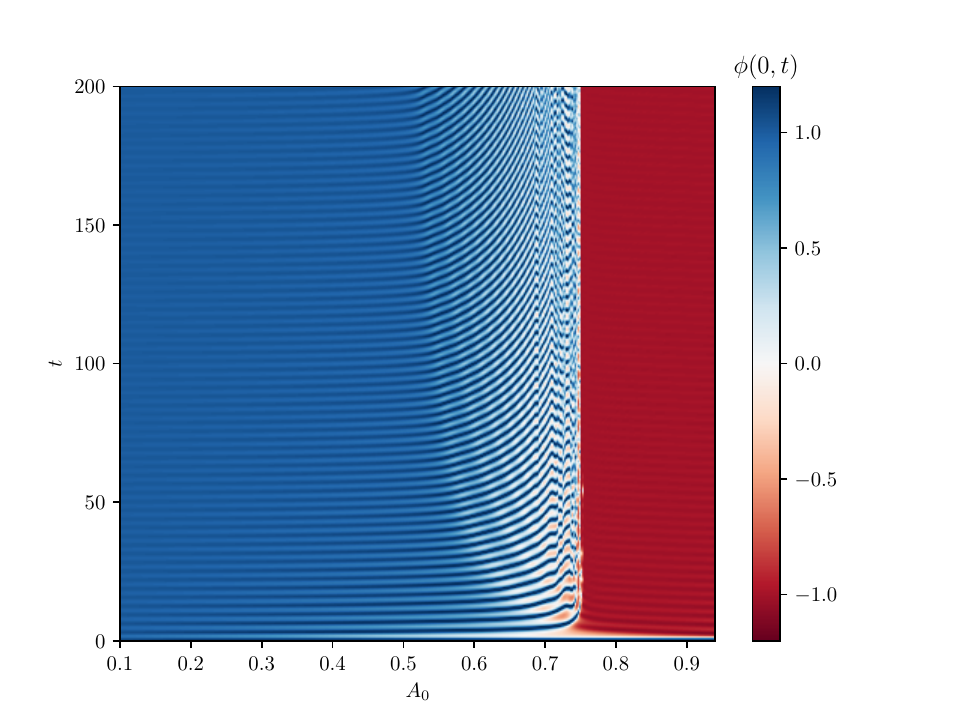}\label{phi_s1}}
    \subfigure[]{\includegraphics[{angle=0,width=8cm,height=6cm}]{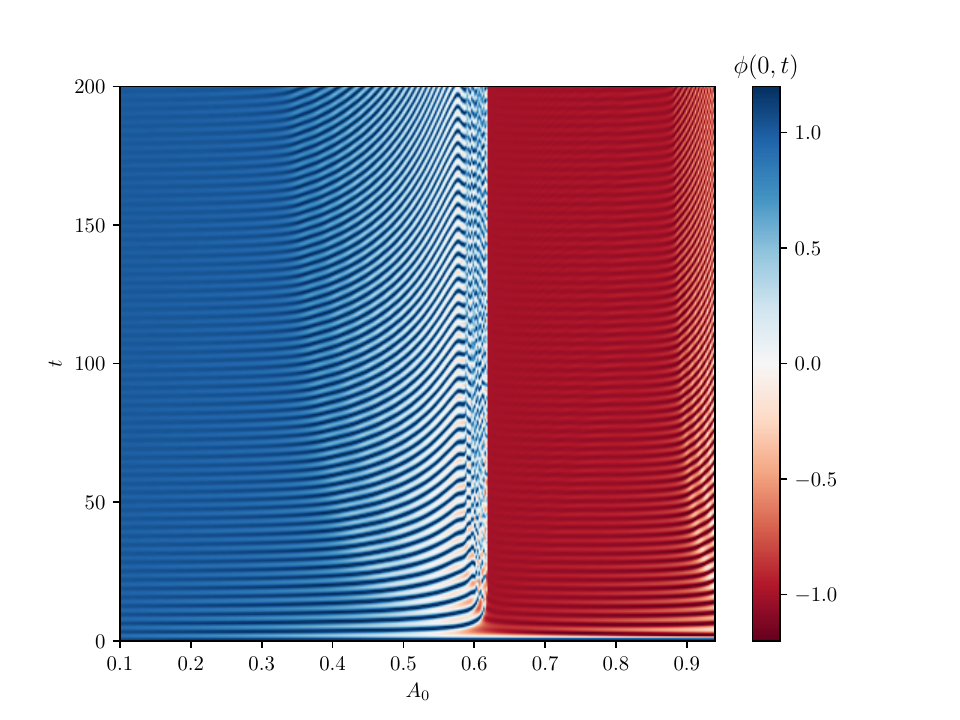}\label{phi_22}}
    \subfigure[]{\includegraphics[{angle=0,width=8cm,height=6cm}]{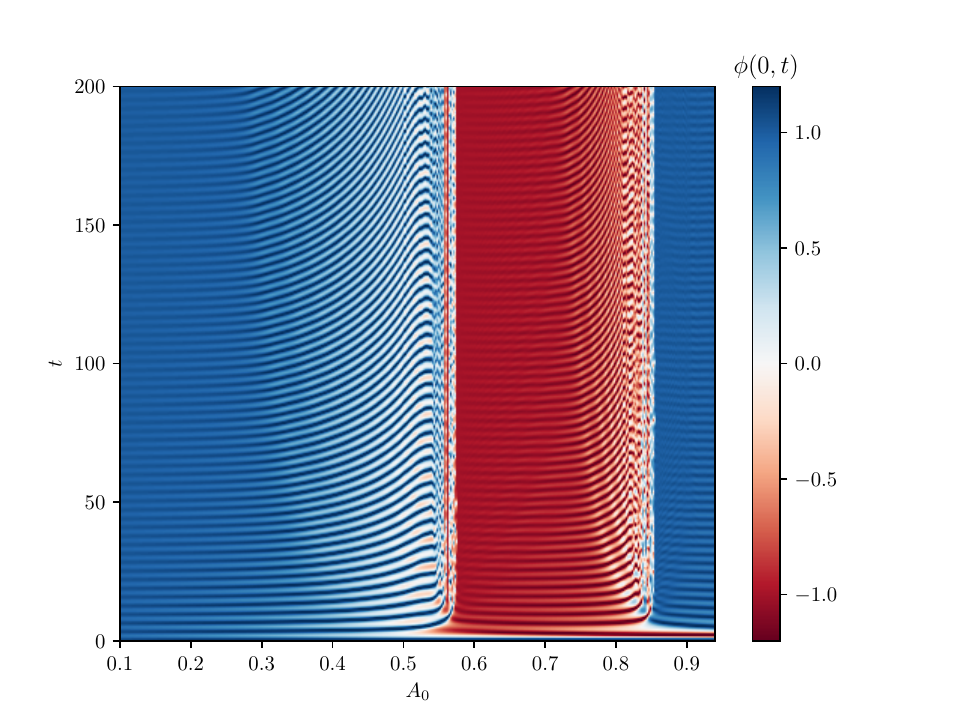}\label{phi_s3}}
    \subfigure[]{\includegraphics[{angle=0,width=8cm,height=6cm}]{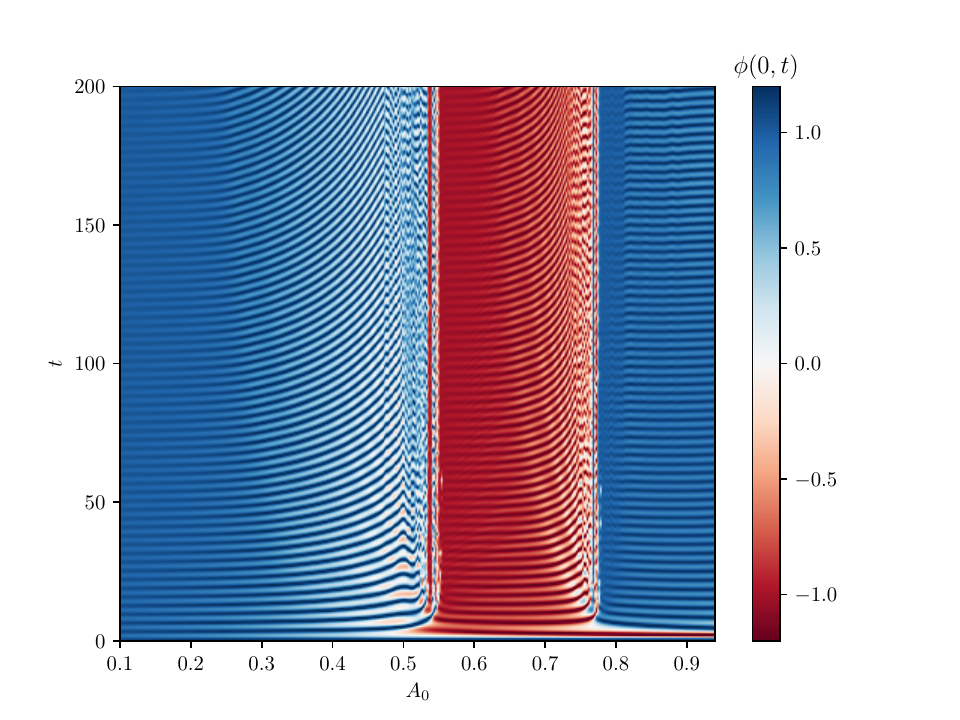}\label{phi_s4}}
  \caption{The resonant structure related to the evolution of the vacuum state $\phi_0=+1$ when perturbed by an oscillon. Again, the blue region means oscillations around that state, while the red band represents an antikink-kink pair at $x=0$. Here, we have fixed (a) $\sigma=1$, (b) $\sigma=3$, (c) $\sigma=5$ and (d) $\sigma=7$.}
  \label{phi_A_sigma}
\end{center}
\end{figure*}

As the amplitude continues to increase, a new blue band appears. It comprises either the formation of a reminiscent oscillon or the emergence of two antikink-kink pairs. For $A_0>0.90$, an intriguing aspect is revealed as this region presents multiple red peaks, see Fig. \ref{tAs10_1}. These peaks suggest the formation of two pairs that scatter symmetrically, with an additional third pair located at $x=0$.

We have also investigated how the values of $\sigma$ (i.e. the width of the initial oscillon) affect the creation of antikink-kink pairs. In this sense, we have explored the evolution of $\phi(x,t)$ for different values of such parameter. Figure~\ref{phi_A_sigma} illustrates the results of the field decay for $\sigma=1,3,5$ and $7$. The blue region again means oscillations around the vacuum state $\phi_0=+1$, while the red color represents an antikink-kink pair at $x=0$. Notice that only two regions — one in blue and one in red — are seen for small $\sigma$. Moreover, as $\sigma$ increases, the width of the red band (i.e. the range of amplitudes) decreases. Interestingly, for large $A_0$, a new blue zone appears as $\sigma$ increases. In contrast to $\sigma=10$, three antikink-kink pairs are not produced under these circumstances. This allows us to comprehend that the rise in $\sigma$ and $A_0$ offers sufficient conditions to perturb the oscillon in order to generate a more complex dynamic.

\begin{figure*}[!ht]
\begin{center}
  \centering
    \subfigure[]{\includegraphics[width=0.3
        \textwidth]{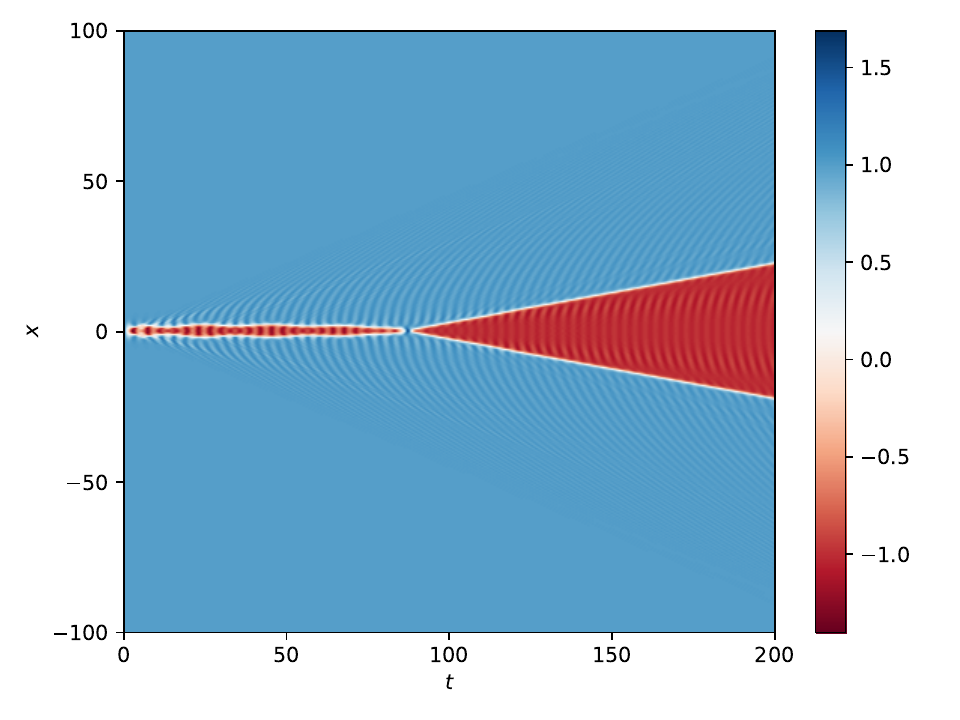}\label{s20A055}}
  \subfigure[]{\includegraphics[width=0.3
    \textwidth]{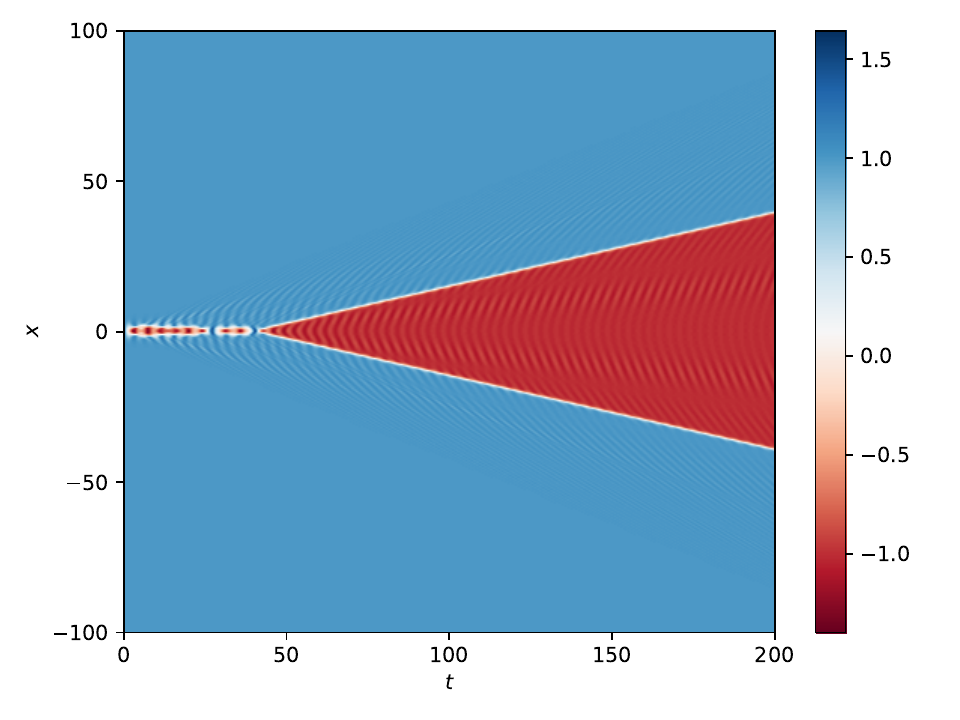}\label{s20A053}}
  \subfigure[]{\includegraphics[width=0.3
    \textwidth]{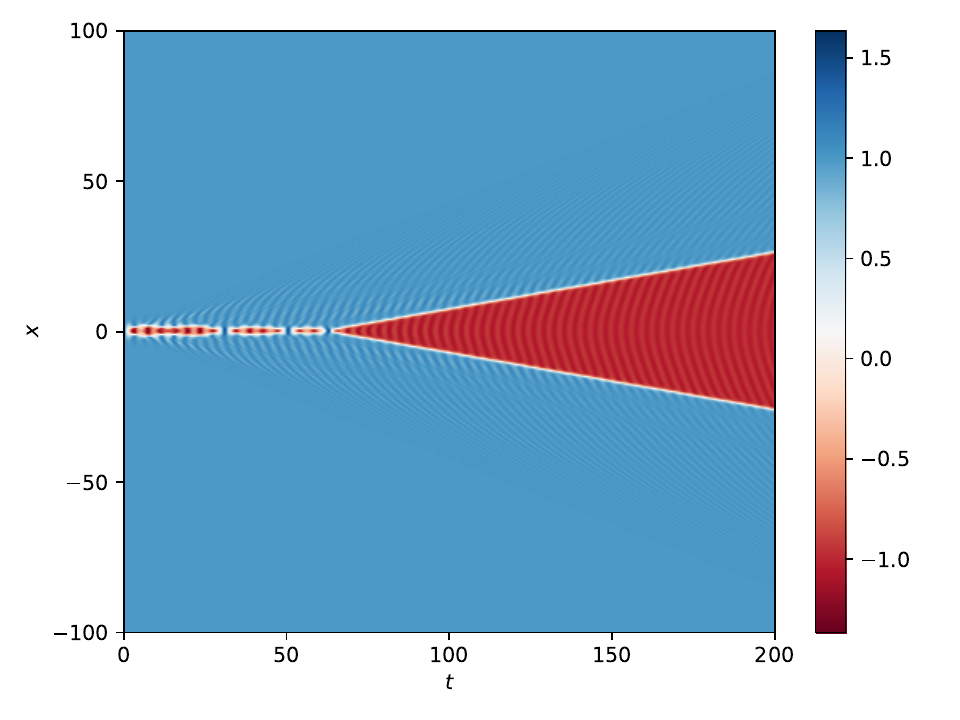}\label{s20A05346}}
   \subfigure[]{\includegraphics[width=0.3
     \textwidth]{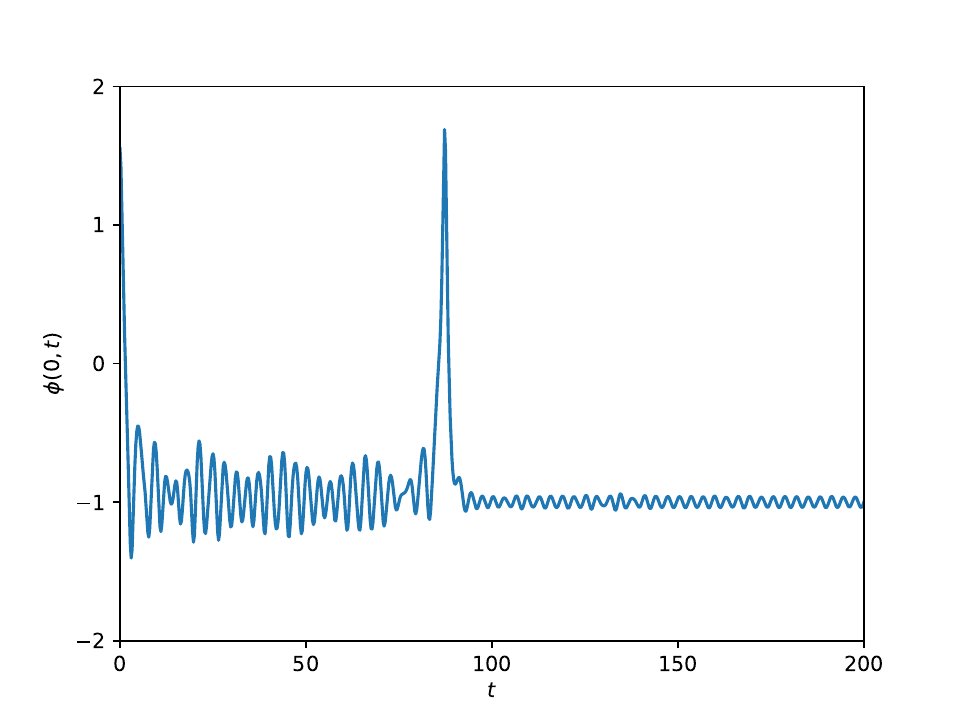}\label{s20A055_1}}
   \subfigure[]{\includegraphics[width=0.3
      \textwidth]{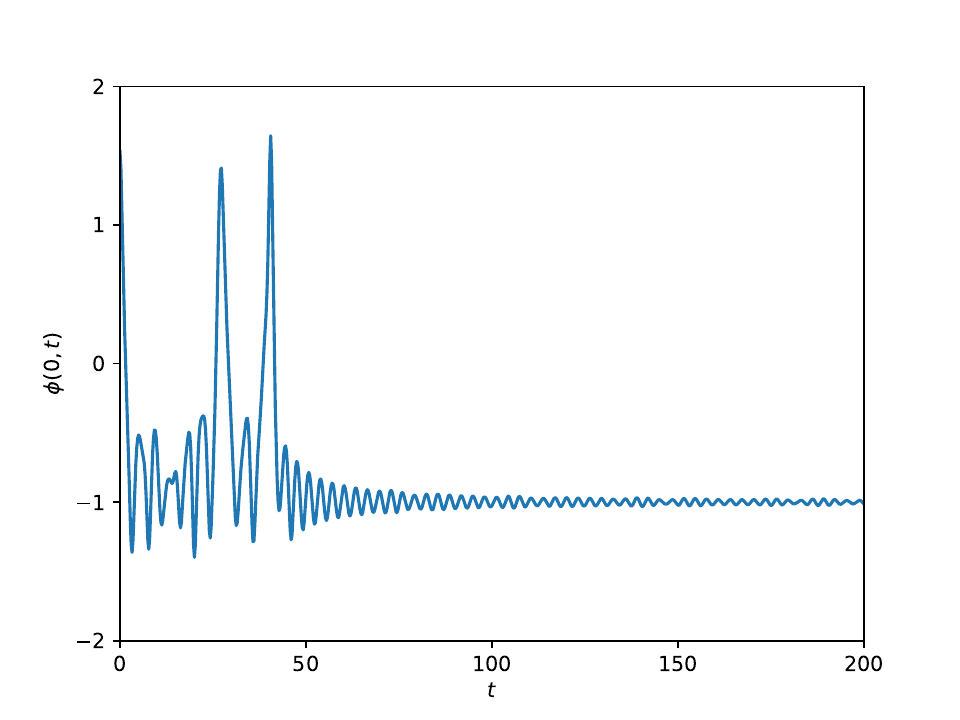}\label{s20A053_1}}
  \subfigure[]{\includegraphics[width=0.3
    \textwidth]{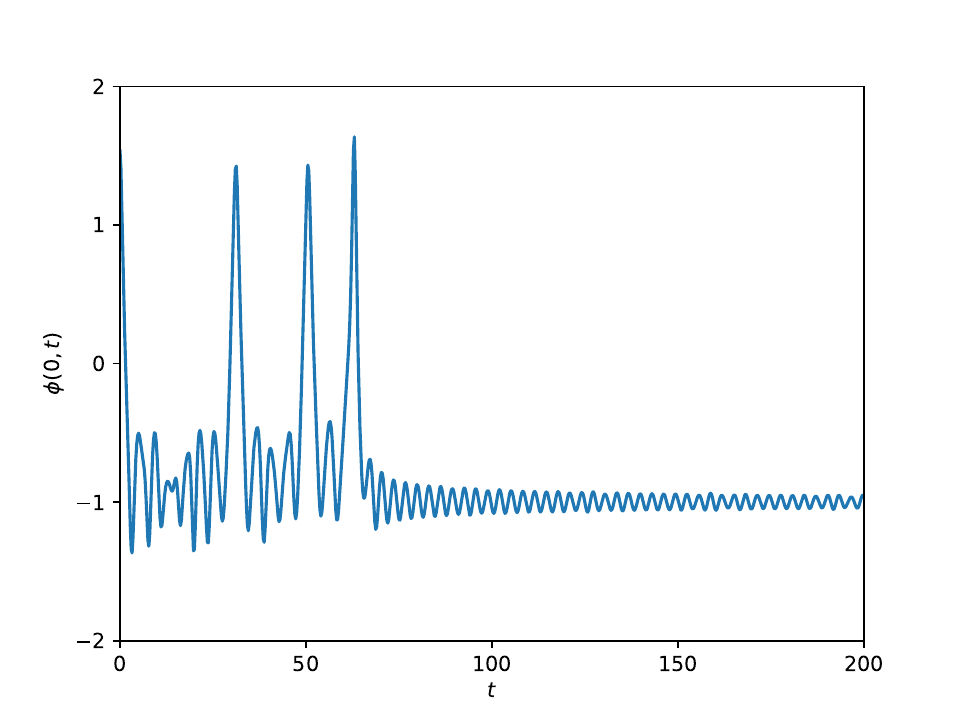}\label{s20A05346_1}}
  \caption{Evolution of $\phi(x,t)$ in the full space (top) and the field profile at the center of mass (bottom). We have chosen $A_0=0.55$ (left), $0.53$ (center) and $0.5345$ (right). Here, $\sigma=20$. The amplitude of the initial oscillon determines the number of bounces performed by the single pair.}
  \label{fig_s20}
\end{center}
\end{figure*}

We also depict the evolution of $\phi(x,t)$ in the full space and the field profile at the center of mass. These results appear in Figure \ref{fig_s20} for $\sigma=20$ and different $A_0$. A single antikink-kink pair is generated as a consequence of the original perturbation. The amplitude of the original oscillon now determines the number of bounces performed by that pair. For instance, when $A_0=0.55$, the initial configuration oscillates and briefly separates. However, it lacks sufficient kinetic energy and hence the pair collides. Only after this first collision, the process of complete separation begins. Interestingly, we see that the pair formation occurs only after two collisions for $A_0=0.53$. In the last case, the solution formed when $A_0=0.5345$ remains trapped for a long time. It performs three complete collisions and only then the antikink-kink pair escapes to infinity.


\section{The initial oscillon as a kink-antikink pair} \label{secIII}


The $\phi^4$ model supports static kinks with particle-like properties. The scattering between these solutions has been intensively studied, see Refs. \cite{campbell,anninos}. In addition, our numerical work has revealed that, once the vacuum state is perturbed by an oscillon, the evolution of $\phi(x,t)$ generates a resonant structure whose pattern mimics that of a genuine kink-antikink collision. This coincidence suggests that the pattern inherent to the field decay is determined by a mechanism similar to the energy exchange that occurs in a kink-antikink interaction.

Therefore, it sounds reasonable to conjecture that the field evolution that comes from the initial condition (\ref{ic}) can be mimicked via a lumplike profile written as a kink-antikink pair.

The mapping of the initial oscillon via well-known solutions was already explored in Ref. \cite{osc_weres}. In that work, however, the authors have approximated their model by the sine-Gordon one. As a consequence, they have mapped the field decay in terms of the usual sine-Gordon breather.

We now explore an additional possibility. Here, we map the initial oscillon via genuine solutions of the original potential (\ref{potential}). These solutions are static objects with particle-like properties. The initial oscillon can be approximated by a static lump constructed as a kink-antikink pair. In this sense, instead of $\phi(x,0) = 1 + A_0 e^{-x^2/\sigma}$, we now adopt the initial condition as
\begin{equation}
\phi \left( x,0\right) =1+\mathcal{A}_{0}\left[ \tanh \left( x+\left\vert
x_{0}\right\vert \right) -\tanh \left( x-\left\vert x_{0}\right\vert \right) %
\right] \text{.}\label{ll}
\end{equation}%
Here, $\mathcal{A}_{0}$ is the amplitude of the lumplike perturbation. In addition, $2x_{0}$ stands for the distance between the kink and the antikink that compose the pair (i.e. the ``width" of the lump whose center is assumed to be at $x=0$).

As we demonstrate below, the general form (\ref{ll}) can be used to map some of the results that arise from the original perturbation (\ref{ic}). The accuracy of the mapping now depends on the parameters inherent to the initial oscillon.

We write down the Taylor series expansions for both (\ref{ic}) and (\ref{ll}). In the sequence, we compare the relevant terms in both series. One then gets that $\mathcal{A}_{0}$ and $x_0$ can be expressed in terms of $A_0$ and $\sigma$ as
\begin{equation}
\mathcal{A}_{0}=\frac{A_{0}}{2\sqrt{1-\frac{1}{\sigma }}}\text{,}
\end{equation}%
and
\begin{equation}
\left\vert x_{0}\right\vert =\text{arctanh}%
\left[ \sqrt{1-\frac{1}{\sigma }}\right] \text{.}
\end{equation}%
Here, we have assumed $\sigma>1$ in order to satisfy $\left\vert x_{0}\right\vert>0$.

We promptly rewrite the initial condition (\ref{ll}) in the form
\begin{equation}
\phi \left( x,0\right) =1+\frac{A_{0}}{2\sqrt{1-\frac{1}{\sigma }}}\left[ \tanh \left( x+\text{arcsech}\left( \sigma ^{-\frac{1}{2}}\right) \right) -\tanh \left( x-\text{arcsech}\left( \sigma ^{-\frac{1}{2}}\right) \right) %
\right] \text{.}\label{app}
\end{equation}%

Equation (\ref{app}) can be used as a good approximation to map some of the results previously obtained. For instance, the solutions for small $A_0$ suggest that the resulting long-lived fluctuations can be treated as a kink-antikink pair that performs multiple bounces. In this interpretation, the individual solutions do not get enough kinetic energy to escape from each other. Due to the nonintegrable nature of the $\phi^4$ model, the colliding pair continuously emits radiation and finally disappears. This forces the field to return to its original vacuum state. This behavior mimics the very same [qualitative] picture that arises when the vacuum configuration is perturbed by an oscillon, see Fig. \ref{fig_s10_A01}.

Naturally, as $A_0$ increases, the correspondence loses accuracy and can not be applied anymore.




Similar restriction was also encountered in Ref. \cite{osc_weres}. There, the mapping based on the sine-Gordon breather only holds for small $A_0$. As this parameter increases, that mapping also loses accuracy. In that work, the authors have not explored the mapping for increasing $A_0$.

In what follows, we adapt our approximation in order to study the mapping for increasing $A_0$. In order to restore accuracy, we assume small values of width $\sigma$. In this case, the initial oscillon is even more localized around $x=0$. As a consequence, the series expansion applies in an even better way.

\begin{figure*}[!ht]
\begin{center}
  \centering
    \subfigure[]{\includegraphics[width=0.47 \textwidth]{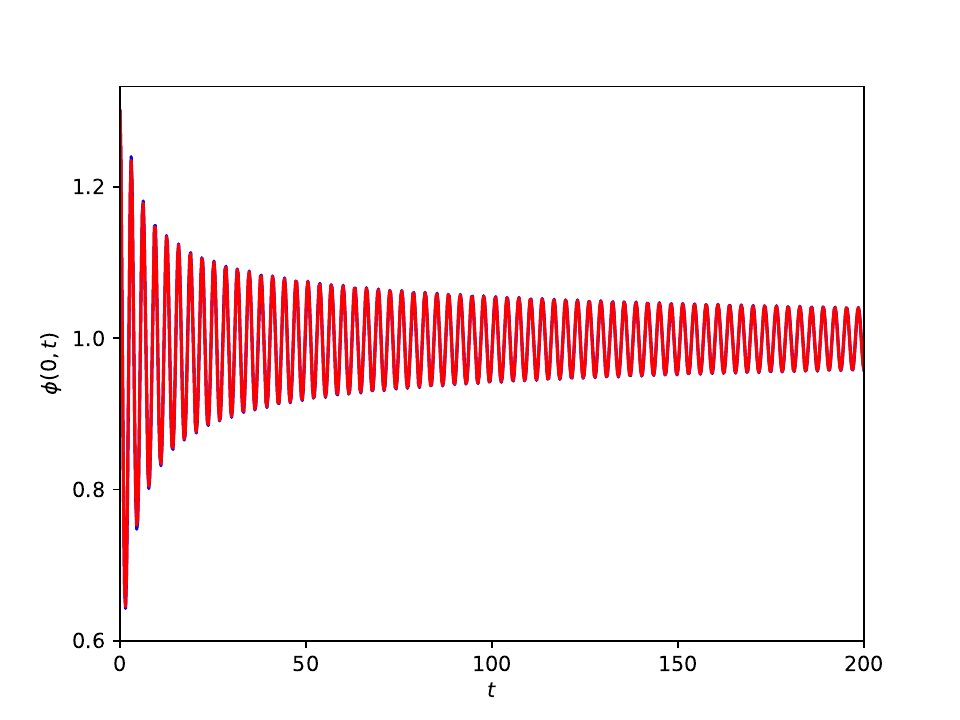}\label{2Ds2_1}}
    \subfigure[]{\includegraphics[width=0.47 \textwidth]{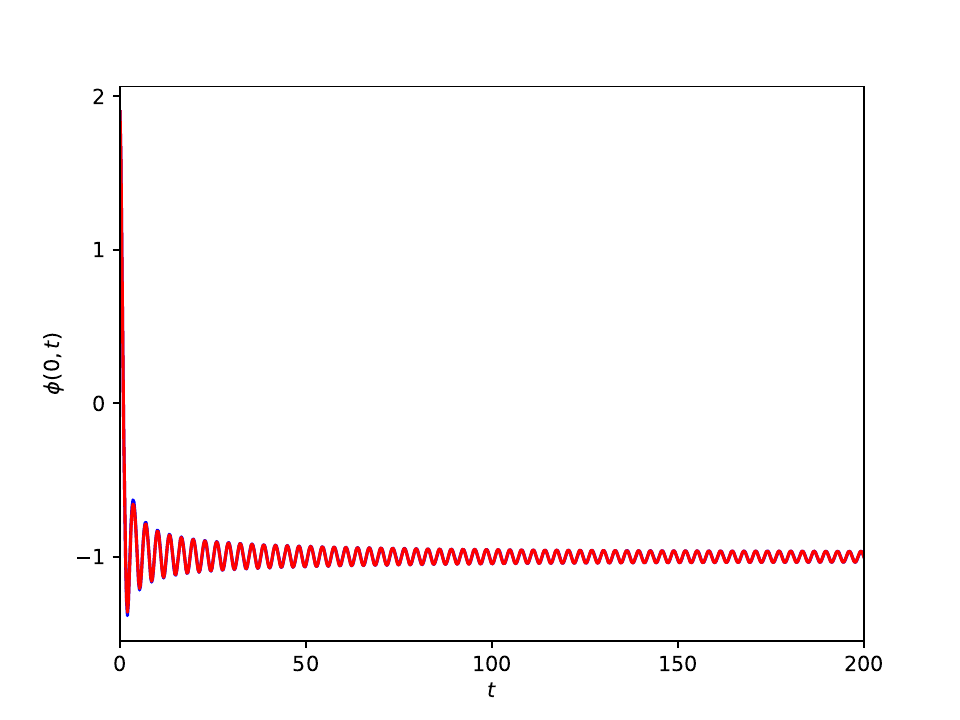}\label{2Ds2_3}}
  \caption{Field profile at the center of mass for $A_0=0.30$ (left) and $A_0=0.90$ (right). Here, $\sigma=2$. Blue line is the solution related to the oscillon (\ref{ic}). Red line represents the result obtained via the lumplike perturbation (\ref{app}). The mapping can now be applied in the limit of increasing $A_0$.}
  \label{2D_sigma_10_B}
\end{center}
\end{figure*}

The results for $\sigma=2$ and $A_0=0.30$ and $0.90$ appear in Figure \ref{2D_sigma_10_B}. It is possible to see how accurate [both qualitatively and quantitatively] the solutions provided by the lumplike perturbation are. However, we reinforce that the final configurations depend on the values of $\sigma$, see the results depicted in Fig. \ref{fig_s20} for $\sigma=20$, for instance. In this sense, we clarify that, when $\sigma=2$, the final configuration for $A_0=0.30$ is a centered short-lived reminiscent oscillon. On the other hand, for $A_0=0.90$, the field decays into a composite structure formed by a single propagating antikink-kink pair and an oscillon around the negative vacuum. For small values of $\sigma$, multiple antikink-kink pairs were not observed.

In all cases above, the initial oscillon was approached by a lumplike profile constructed as a static kink-antikink pair. The results obtained from Eq. (\ref{ic}) were then mimicked via those related to a kink-antikink interaction. This mapping may contribute to the comprehension on the role played by the energy exchange during the formation of the final configurations. Moreover, the approximation based on the lumplike perturbation may provide some insight about the structure that resembles that of a kink-antikink collision.


\section{Summary and perspectives} \label{secIV}


We have studied the evolution of a vacuum state when perturbed by an oscillon.

We have considered a model with a single scalar field $\phi(x,t)$ and whose potential has the usual $\phi^4$ form. The resulting nonintegrable theory supports static kink solutions with particle-like properties. These configurations admit both translational and vibrational modes. These modes are commonly applied to explain the evolution of a kink-antikink scattering. As a result of the energy exchange between those modes, a resonant structure with a peculiar fractal pattern appears.

Motivated by the important results presented in Ref. \cite{osc_weres}, we have considered that the vacuum state $\phi_0 = +1$ is perturbed by an oscillon. This initial oscillon is defined by two parameters, i.e. $A_0$ and $\sigma$. These values represent, respectively, the amplitude and width of the perturbation.

We have then studied how different values of $A_0$ give rise to different final configurations. In this sense, we have considered fixed values for $\sigma$.

Initially, we have investigated the field decay for small $A_0$ (i.e. a weakly excited oscillon). The results have revealed that a centered reminiscent oscillon around the vacuum state appears. Due to the non-integrability of the $\phi^4$ theory, that structure continuously emits radiation and finally annihilates. The field then returns to its fundamental state.

In the sequence, we have explored the evolution for not so small $A_0$. This range was not considered in Ref. \cite{osc_weres}. Here, as $A_0$ increases, the initial oscillon becomes more and more excited. In this context, field evolution has revealed new interesting configurations. In addition, for intermediary amplitudes, long-lived reminiscent oscillons emerge. Again, they irradiate and, after an extremely long time, annihilate. 

We have found that, as $A_0$ approaches $1$, composite solutions emerge. These configurations are novel and find no correspondence with the ones introduced in \cite{osc_weres}. Here, the composite profiles are formed by antikink-kink pairs and centered oscillons. For $A_0=0.60$ and $A_0=0.80$, one and two pairs emerge, respectively. When $A_0=0.95$ (i.e. a highly perturbed initial oscillon), three kink-antikink pairs arise. In all cases, centered oscillons appear. For $A_0=0.60$ and $A_0=0.95$, the field oscillates around the state $\phi_0=-1$. On the other hand, when $A_0=0.80$, the fluctuations occur around $\phi_0=+1$.

We have also investigated the field decay for larger $\sigma$. In this case, we have fixed $\sigma=20$. We have observed that a single antikink-kink pair is generated. The amplitude $A_0$ now determines the number of bounces performed by the pair. These multiple bounces can be labeled as usual, i.e. based on the order of the window they belong. The existence of windows with nontrivial order is currently understood as a consequence of the energy exchange between the internal modes.

Motivated by these results, we have obtained the corresponding resonant structure. Notably, it has revealed a pattern that mimics that of a genuine kink-antikink scattering.

In view of this coincidence, we have mapped the initial oscillon as a lumplike profile constructed as a kink-antikink pair. The numerical simulations have revealed that the mapping holds well for small $A_0$. However, it naturally loses accuracy as $A_0$ increases. Similar limitation was also encountered in Ref. \cite{osc_weres}, where the authors have proposed a map based on the sine-Gordon breather.

Moreover, the authors of Ref. \cite{osc_weres} have not explored the mapping for increasing $A_0$. In order to restore accuracy and map the evolution for larger $A_0$, we have assumed small $\sigma$. In this case, the initial oscillon is even more localized around $x=0$.

The results for $\sigma=2$ have demonstrated how accurate the new approximate solutions are for different $A_0$. In particular, when $A_0=0.90$, we have verified that the field again decays into a composite structure.

This successful mapping may contribute to the understanding of the role played by a mechanism that mimics the energy exchange during the decay process, the formation of the resonant structure.

Future investigations include the application of the present idea to the $\phi^6$ and the sine-Gordon potentials. The last one sounds particularly interesting due to the integrable nature of that theory. We plan to study the resonant structures and verify whether the pattern can be compared to some well-established one. On the other hand, this pattern can always reveal a novel profile to be explored. These issues are currently under investigation, and positive results will be eventually reported in a future contribution.


\section*{Acknowledgements}

The authors thank João Campos and Azadeh Mohammadi for their useful comments. EH also thanks the partial financial support received from the Conselho Nacional de Desenvolvimento Científico e Tecnológico - CNPq (Brazillian agency) via the Grant N° 309604/2020-6. This study was financed in part by the Coordenação de Aperfeiçoamento de Pessoal de Nível Superior - Brasil (CAPES) - Finance Code 001.



\end{document}